\documentclass[onecolumn,superscriptaddress,aps,preprintnumbers,amsmath,amssymb]{revtex4-1}
\usepackage{graphicx}
\usepackage{dcolumn}
\usepackage{bm}
\usepackage{algorithm2e}
\usepackage{lipsum}
\usepackage{xfrac}
\renewcommand{\figurename}{Fig.}
\renewcommand\tablename{Table}
\begin{document} 
\title{Manuscript Title:\\with Forced Linebreak}
\author{Chumki Nayak}
\affiliation{Department of Physical Sciences, Bose Institute, 93/1, Acharya Prafulla Chandra Road, Kolkata 700009, India}
\author{Suvadip Masanta}
\affiliation{Department of Physical Sciences, Bose Institute, 93/1, Acharya Prafulla Chandra Road, Kolkata 700009, India}
\author{Sukanya Ghosh}
\affiliation{Institute for Condensed Matter and Complex Systems, School of Physics and Astronomy, The University of Edinburgh, Edinburgh, EH9 3FD}
\author{Shubhadip Moulick}
\affiliation{Department of Condensed Matter and Materials Physics, S.N. Bose National Centre for Basic Sciences, Kolkata 700106, India}
\author{Atindra Nath Pal}
\affiliation{Department of Condensed Matter and Materials Physics, S.N. Bose National Centre for Basic Sciences, Kolkata 700106, India}
\author{Indrani Bose}
\affiliation{Department of Physical Sciences, Bose Institute, 93/1, Acharya Prafulla Chandra Road, Kolkata 700009, India}
\author{Achintya Singha}
\email{achintya@jcbose.ac.in}
\affiliation{Department of Physical Sciences, Bose Institute, 93/1, Acharya Prafulla Chandra Road, Kolkata 700009, India}

\begin{abstract}
Monolayer transition metal dichalcogenides (TMDCs) constitute the core group of materials in the emerging semiconductor technology of valleytronics. While the coupled spin-valley physics of pristine TMDC materials and their heterstructures has been extensively investigated, less attention was given to TMDC alloys, which could be useful in optoelectronic applications due to the tunability of their band gaps. We report here our experimental investigations of the spin-valley physics of the monolayer and bilayer TMDC alloy, MoS$_{2x}$Se$_{2(1-x)}$, in terms of valley polarization and the generation as well as electrical control of a photocurrent utilising the circular photogalvanic effect. Piezoelectric force microscopy provides evidence for an internal electric field perpendicular to the alloy layer, thus breaking the out-of-plane mirror symmetry. The experimental observation is
supported by first principles calculations based on the density functional theory. A comparison of the photocurrent device, based on the alloy material, is made with similar devices involving other TMDC materials. 
\end{abstract}
\title{ 
Valley polarization and photocurrent generation in transition metal dichalcogenide alloy MoS$_{2x}$Se$_{2(1-x)}$}
\maketitle

\subsection*{Introduction}
The two-dimensional (2D) transition metal dichalcogenides (TMDCs) with the chemical formula, MX$_2$, where M $=$ Mo, W and X $=$ S, Se, Te are a transition metal and a chalcogen atom respectively, have drawn significant attention from researchers in recent years. This is because of their immense application potential in semiconductor-based technologies like valleytronics, which exploits the valley degree of freedom (DOF) besides those of spin and charge \cite{Schaibley, Xiaodong, Mak, Lu}. The combination of time-reversal symmetry (TRS), broken inversion symmetry (IS) and the presence of significant spin-orbit interactions (SOI) involving the “heavy” $d$ electrons of the transition metal ions gives rise to spin-valley locking and valley-specific physical properties and phenomena. Figure 1a and b shows the coexistence of a pair of valence band (VB) and conduction band (CB) valleys in the band structure located at two inequivalent points K and K$^\prime$ in momentum space with the K and  K$^\prime$ valleys related to each other via TRS. The Zeeman-type splitting of the spin-degenerate VB and CB, a consequence of broken in-plane IS and SOI, is shown with the split subbands designated as VB-, VB+, CB- and CB+ in the order of increasing energy. The splitting in the case of the VB is an order of magnitude larger than that of the CB. The split VB subbands are spin-polarised, with the electron spins oriented in the out-of-plane direction (up or down). Due to TRS, the polarizations of the subbands are reversed at the K and K$^\prime$ valleys. This gives rise to the unique feature of spin-valley locking in the monolayer implying that the spin and valley degrees of freedom are coupled \cite{Schaibley, Xiaodong, Ziao}.

Focusing on the TMDC monolayer material MoX$_2$, there are two other distinct classes of related monolayer materials: the Janus TMDC materials, MoXY (X, Y$=$S, Se, X$\neq$Y) and the TMDC alloy (TMDCA), MoS$_{2x}$Se$_{2(1-x)}$ $(0\le x\le 1)$. The three classes of monolayer  materials have a common structural ingredient, namely, a central layer of Mo ions. This layer is sandwiched between two layers of chalcogen atoms. In the monolayer Janus MoSSe, one chalcogen layer is occupied by S atoms and the other layer by Se atoms, thus breaking the out-of-plane mirror symmetry. The polar nature of the material generates an electric dipole moment with its associated electric field perpendicular to the monolayer. This intrinsic electric field generates the Rashba SOI which splits the VB at the $\Gamma$ point in momemtum space. In the monolayer TMDCA, the chalcogen layers are randomly decorated with the S and Se atoms (Fig. 1c) introducing disorder in the system. In this context, two  pertinent questions  are whether the universal features of the spin-valley physics of pristine TMDC materials survive in the presence of disorder and whether an intrinsically generated out-of-plane electric field is present in the alloy, as in the case of the Janus material, or disorder averages the field to be zero. From an application point of view, the issues are tied to the possibility of constructing photocurrent devices based on monolayer and bilayer TMDCA materials.

In the cases of multilayer and bulk TMDC materials, the band-gap is indirect and the IS is not broken when the number of layers is even. The breaking of the IS in bilayer TMDC materials, an essential requirement for applications involving valley-contrasting properties, is achieved by applying an external electric field perpendicular to the bilayer \cite{DuL}, resulting in a potential difference between the layers. In this context, the Janus TMDC materials are of special note as the IS is intrinsically broken for both the odd and even multilayer materials thus bypassing the necessity of using an external electric field \cite{LuM, Zhang}. In our study, we investigated the spin-valley physics of the monolayer and bilayer TMDCA through experiments as well as first principles calculations based on density functional theory (DFT). We carried out polarization-resolved photoluminescence (PL) and photocurrent generation experiments using off- and near on-resonance lasers with energies of 2.54 eV and 1.96 eV, respectively. In the case of the off-resonance laser excitation, the experimental measurements of the degrees of valley and photocurrent polarization of monolayer TMDCA yielded some surprising features. The optical and electrical control of the photocurrent was brought about by changing the polarization of the laser light and applying an external electric field via a gate voltage. The results obtained clearly establish the coceptual validity of the spin-valley physics in the presence of disorder in the chalcogen layers of the alloy material. The result is not surprising as the predominant contributions to the valley physics come from the $d$ electrons of the Mo atoms located in the central layer, which is the common feature in all three classes of TMDC materials.  The disorder in the chalcogen layers could in principle yield a zero average out-of-plane electric dipole moment and field. To check this, piezoelectric force microscopy (PFM)  was used which provides clear evidence for the existence of a finite  internal electric field perpendicular to the TMDCA monolayer. The photocurrent device based on the monolayer/bilayer alloy material was shown to compare quite well, performance-wise, and in certain respects better, with similar devices involving other TMDC materials. Our study provides the first experimental demonstration of the generation of a photocurrent in monolayer and bilayer TMDCA samples. Earlier experimental studies \cite{Eginligil, Quereda, Yuan, Cha, Liu} were focused on  monolayer/bilayer/bulk TMDC materials with no study reported so far on the Janus material MoSSe or the related class of TMDCA materials.
\begin{figure*}[t!]
 \centering
\includegraphics[width=0.9\linewidth]{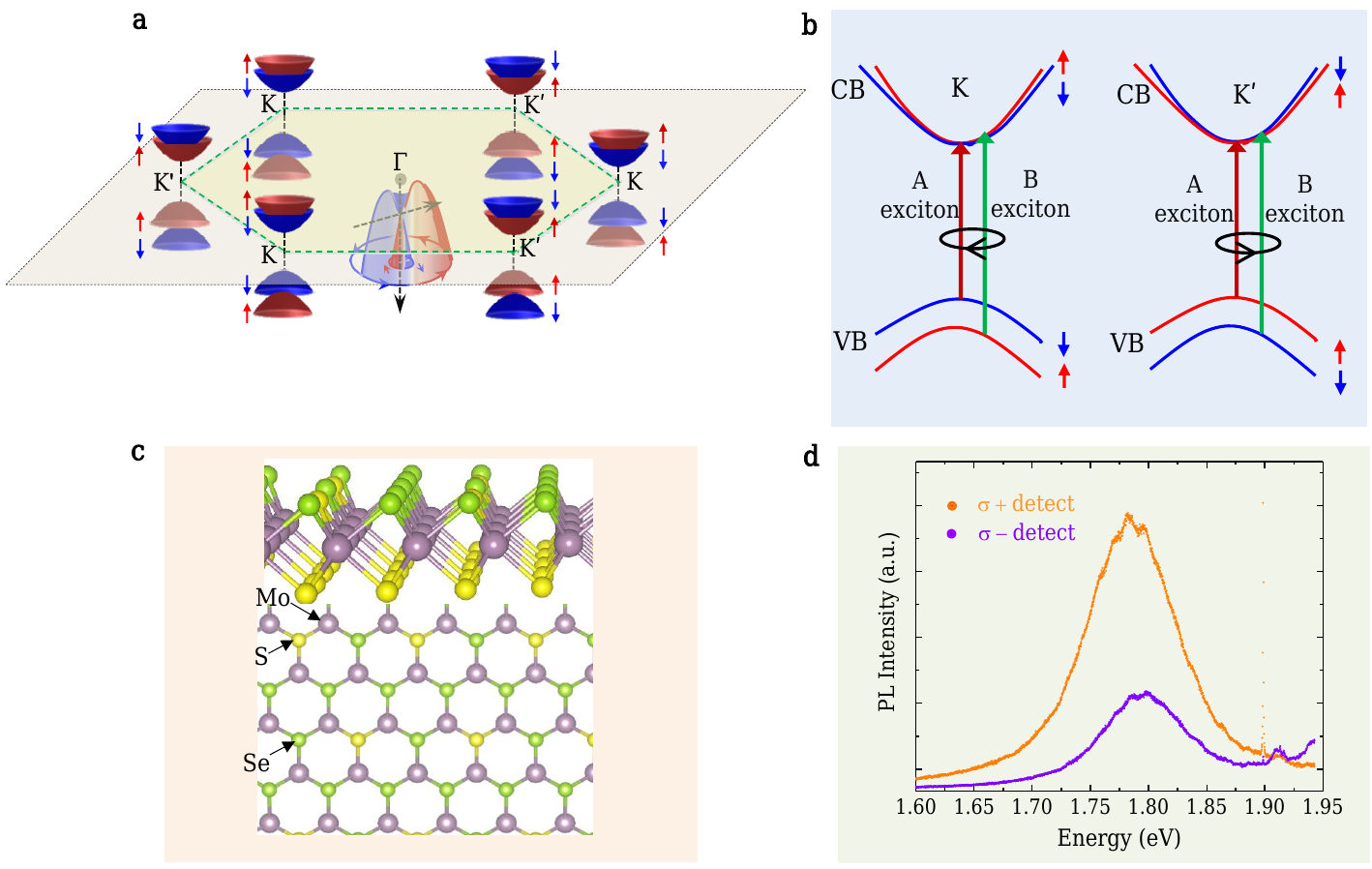}
\caption{Schematic representation of Brillouin zone, band diagram of K, K$^{\prime}$ valleys, crystal structure and PL of monolayer TMDCA. {\bf{a}} Schematic representation of the valley structures at the  K, K$^{\prime}$ and $\Gamma$ points of Brillouin zone. {\bf{b}} Interband transitions at K and K$^{\prime}$ points under left and right circularly polarized light resulting in  the formations of  A and B excitons. {\bf{c}} Side (upper panel) and top (lower panel) views of monolayer TMDCA structure.  {\bf{d}} Polarization-dependent PL measurement under near on-resonance 1.96 eV laser excitation energy.}
\label{Fig. 1.}
\end{figure*}
\subsection*{Results}
\subsubsection*{First principles calculations}
 The k.p. model derivations and first principles calculations are summarized in Supplementary Note 1. Figure 2 shows the fully-relativistic band structure of the TMDCA alloy (x$=$0.6) including spin-orbit coupling, along the high-symmetry directions in the Brillouin zone. The highest VB and the lowest CB spin splittings at the K, K$^{\prime}$ valleys have magnitudes $\sim$ 170 meV and $\sim$ 10 meV respectively (see Supplementary Fig. 1). The band structure for the alloy material is colour-coded for the spin polarization component $<s_z>$. The expectation value of the spin components corresponding to the $i$-th-energy spinor eigenfunction, $\Psi_{i, {\bf k}}({\bf r})$, are obtained as: 
\begin{equation}
<s_{\alpha}> = \frac{1}{2} \frac{<\Psi_{i,k}\vert{\sigma_{\alpha}} \vert \Psi_{i,k}>}{<\Psi_{i,k}\vert \Psi_{i,k}>} 
\end{equation}
where ${\sigma_{\alpha}}$, $\alpha=x, y, z$, are the Pauli spin matrices. The band structure confirms the universal features of the spin-valley physics that the alloy material shares with Janus and TMDC monolayers \cite{Wang, Yang, Yu, Hu}, schematized in Fig. 1a and b. In particular, the K and K$^{\prime}$ valleys exhibit opposite polarization values, $<s_{z}> = {\pm} 1/2$, with the computed sequence of values for the split VB and CB bands as shown in Fig. 1b. We carried out a PFM experiment to establish the presence of an intrinsic electric field  $E_\mathrm{int}$  perpendicular to the TMDCA monolayer. The field is associated with the breaking of the out-of-plane mirror symmetry. The Rashba SOI is generated by $E_\mathrm{int}$. The splitting of the VB at the $\Gamma$ point (inset of Fig. 2) and in-plane spin texture (Supplementary Fig. 4a) observed in the first principles calculations are consistent with the experimental result of a non-zero $E_\mathrm{int}$. No Rashba splitting is seen in the pristine MoX$_2$ systems as the electric field $E_\mathrm{int} = $ 0 in these systems (see Supplementary Fig. 5). 
\begin{figure}
 \centering
\includegraphics[width=0.5\linewidth]{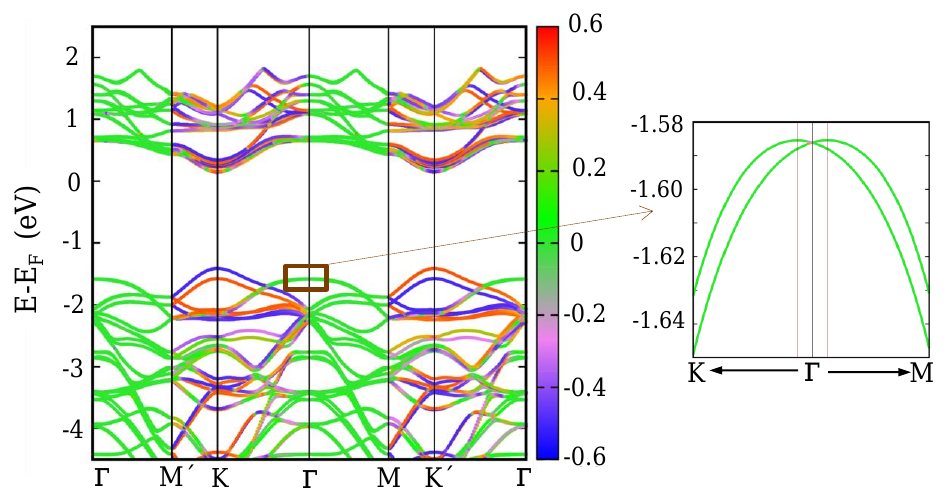}
\caption{ DFT calculated electronic band structure of monolayer TMDCA. It shows the electronic band structure for MoSSe alloy projected on $<s_z>$. The inset shows the magnified Rashba splitting close to the $\Gamma$ point.}
\label{Fig. 1.}
\end{figure}
\subsection*{Experimental Results}
\subsubsection*{Sample characterizations}
The detailed sample preparation method using chemical vapor deposition (CVD) synthesis procedure is discussed in the method section. Triangular-shaped flakes of TMDCA are observed randomly on the substrates, as confirmed by optical imaging (Supplementary Fig. 7b). A scanning electron microscopy (SEM) image of a single TMDCA flake is shown in Supplementary Fig. 7c. To determine the atomic percentage of Mo, S, and Se in the TMDCA, energy-dispersive X-ray spectroscopy (EDX) was performed. The EDX spectrum and estimated percentages of Mo, S and Se from this study are given in Supplementary Fig. 7d. The atomic force microscopy (AFM) image and height profile of the TMDCA sample (see Supplementary Fig. 8a) indicates that its thickness is in the monolayer regime. The PL spectrum (Supplementary Fig. 8b) exhibits the components resulting from the recombination of the A and B excitons (X$^0$) and also the A trion. The energy associated with the A trion, A exciton, and B exciton is determined to be 1.78 eV, 1.81 eV, and 1.95 eV, respectively. Supplementary Fig. 8c shows the Raman spectrum of the TMDCA sample, with the most prominent Raman modes observed being E$^1_1$, A$^1_1$, E$^3_2$, E$^2_1$ and A$^2_1$ at 231 cm$^{-1}$, 272 cm$^{-1}$, 368 cm$^{-1}$, 384 cm$^{-1}$ and 405 cm$^{-1}$ respectively. These observed modes are consistent with previous reports \cite{Bera, Masanta, Mukherjee}. The optical image of the phototransistor made of monolayer TMDCA using the technique mentioned in the device fabrication section is presented in Supplementary Fig. 8d. The plot of photocurrent vs light excitation power density with source-drain bias, V$_{ds} =$ 1V, is depicted in Supplementary Fig. 8e showing a linear dependence on illumination intensity. All subsequent photocurrent measurements were conducted in this linear region. Supplementary Fig. 8f displays the transfer characteristics of the TMDCA phototransistor at V$_{ds} =$2V under both dark and illuminated conditions using an excitation wavelength of 488 nm. The photocurrent shows a slight increase when the device is in the OFF state (V$_g < $10V). However, at V$_g >$ 10V, there is a rapid enhancement in the photocurrent. In this ON state, the current contribution arises not only from a significant increase in photogenerated carriers but also from carrier tunneling. All the subsequent measurements have been performed in the OFF state.
\subsubsection*{Piezoelectricity in TMDCA}
To experimentally probe the existence of an out-of-plane dipole moment arising due to the out-of-plane asymmetry in the structure of our TMDCA samples, we performed PFM measurements on various samples with different thicknesses. The details and schematic representation of the PFM measurements are provided in Supplementary Note 4 and Supplementary Fig. 9. The PFM topography and height profile shown in Supplementary Fig. 10a and c, indicate that the measured sample is a monolayer. The lack of sharpness in the image can be attributed to the thickness of the monolayer sample \cite{XieS, XueF}. Nonetheless, the piezoelectric contrast between the TMDCA sample and SiO$_{2}/$Si substrate (Supplementary Fig. 10b), and the piezoresponse profile (Supplementary Fig. 10d) along the indicated line in Supplementary Fig. 10b clarify that the observed piezoelectric response originates from the sample.

The operational principle of the PFM is based on the inverse-piezoelectric effect in which an electric field induces a mechanical strain response. The field is generated by applying a voltage between the tip of an AFM and the surface of the sample. The strain response is in terms of surface deformation, either up or down, in addition to the pre-existing topography of the sample surface. The amplitude and the phase of the deformation, in the forms of a butterfly-shaped amplitude switching loop  and nearly 180$^{\circ}$ phase reversal loop under applied bias voltage (Supplementary Fig. 11a and b), confirm the existence of the inherent out-of-plane electric dipole moment (polarization) of the TMDCA sample \cite{XieS, XueF, LuM, Rao}. The corresponding piezoelectric coefficient $d_{33}$ is determined as 0.13$\pm$0.008 pm/V, which is comparable to that of the Janus MoSSe sample \cite{LuM}. In this context, it is pertinent to point out that in the case of the randomized MoSSe alloy, experiments [ref 7] indicate a negligible out-of-plane electric dipole moment. To gain further insight, we chose bilayer and trilayer samples shown in Supplementary Fig. 12a and c. Here, a clear piezoelectric amplitude contrast between the sample and substrate is visible (Supplementary Fig. 12b). The line profile depicts a noticeable change in the piezoelectric amplitude for the bilayer and trilayer compared to the substrate (Supplementary Fig. 12d). The piezoresponse amplitude–voltage butterfly loop and phase–voltage hysteresis loop are presented in Supplementary Fig. 13a and b.
It is noteworthy that with an increasing layer number, the piezoelectric amplitude is enhanced by the strong internal electric filed originating from the parallel electric dipoles of the layers. This observation proves that the measured vertical polarization is an intrinsic material property, not stemming from any topological defects \cite{XueF, LuM}. Additionally, the PFM study of the SiO$_{2}/$Si substrate (Supplementary Fig. 14) does not show any significant piezoresponse, revealing that the piezoelectric behaviour solely originates from alloy TMDCA. The $d_{33}$ value for the trilayer TMDCA reaches 0.17$\pm$0.009 pm/V. It should be noted that the $d_{33}$ values are qualitative, as the effective piezoelectricity of  TMDCA may be sensitive to small variations in electrical properties \cite{LuM}.

\subsubsection*{Valley polarization and photocurrent}
The photoresponse to  circularly polarized light (CPL) is governed by valley-dependent optical selection rules due to which the interband optical transitions at the K and K$^{\prime}$ valleys are driven solely by right circularly polarized (RCP) and left circularly polarized (LCP) light respectively. This differential absorption of circularly polarised light, termed circular dichroism (CD), results in the selective occupation of a CB valley by photoexcited electrons. Due to strong Coulomb interactions,  the photoexcited electron in the CB and the hole in the VB bind each other to form an exciton. In the case of monolayer MoS$_2$, polarised PL experiments provide evidence for two types of excitons: A and B \cite{Liu}. The optical transitions for the two types of excitons in the TMDCA material are indicated in Fig. 1b. These excitons are charge neutral and can further lower their energy by capturing an excess electron or hole to form a negatively or positively charged trion. As detailed below, the valley polarization experiment carried out on the TMDCA sample provides evidence for photogenerated A and B excitons, consistent with the scheme depicted in Fig. 1b.

The exciton-associated valley polarization, achieved via optical pumping, was initially demonstrated for monolayer MoS$_2$ using polarization-resolved PL \cite{Liu}. The exciton has a finite lifetime due to the recombination of the constituent electron and hole. The recombination is accompanied by a characteristic PL which is expected to have the same circular polarization as that of the incident light if the exciton recombination primarily occurs in the same valley in which it is generated. From the experimental data, the degree of valley polarization $\eta$ can be computed as \cite{Schaibley, Shan},
\begin{equation}
\eta = \frac{PL(\sigma+)-PL(\sigma-)}{PL(\sigma+)+PL(\sigma-)}
\end{equation}
where  $PL(\sigma+)$ and $PL(\sigma-)$ are the RCP and LCP components of the emitted PL intensity. The polarization degree is an indicator of the effectiveness of the optical selection rules as well as the preservation of the identity of the valley charge carriers before recombination. Fig. 1d shows the PL intensity versus emission energy of the TMDCA sample with the incident light being right circularly polarized. The light energy is 1.96 eV near on-resonance with the A exciton. The dominant contribution to the PL intensity, with peak position around 1.8 eV, comes from the $PL(\sigma+)$ component and the estimated  $\eta$ value is 42.7$\%$. As the PL spectrum (not polarization-resolved) of Supplementary Fig. 8b shows, the PL intensity is mostly due to the A exciton. The deviation from the ideal value of $\eta=$100$\%$ can be understood in terms of the exciton valley depolarization via intervalley electron-hole exchange interaction and phonon-mediated intervalley scattering at room temperature \cite{Schaibley}. For the monolayer MoS$_2$, a wide range of values for $\eta$ has been reported, from 100$\%$ to 32$\%$ depending on experimental conditions (for example temperature) and sample preparation \cite{Kioseoglu}.

Due to spin-valley locking and valley-specific selection rules for optical excitations, the photoexcited electrons in the CB are both valley- and spin-polarized. In the presence of local electric fields, the excitons dissociate into free carriers which are accelerated by a source-drain voltage to generate a spin-polarized photocurrent. Since intervalley scattering requires spin-flip as well as momentum conservation, the generation of a quite robust spin-polarized photocurrent is possible. The detection of spin polarization in TMDC samples has been reported using appropriate device setups, e.g., a lateral spin-valve structure with ferromagnetic contacts\cite{Cha}. In our experiment, we detect only the charge photocurrent and the data enable us to obtain a quantitative estimate of the polarization degree of the photocurrent. The experimental setup for the generation of a photocurrent in response to polarized light is shown in Fig. 3a. In this setup, the TMDCA is illuminated with light propagating in the $x-z$ plane at an angle of incidence ($\theta$) with respect to the $z$-direction which is normal to the sample plane. A quarter-wave plate (QWP) is used to change the polarization of the light (linearly polarized) incident on it by rotating the plate to change the angle $\varphi$ between the fast axis of the QWP and the incident light polarization. The degree of circular polarization of the light falling on the sample is given by $P_{circ} = sin2\varphi$. As $\varphi$ is changed over a 180$^0$ period, the light polarization successively changes with linear polarization at the angles $\varphi= $ 0$^0$, 90$^0$, 180$^0$ $(P_{circ}=0)$, RCP ($\sigma+$) at $\varphi= $ 45$^0$ $(P_{circ}=1)$ and LCP ($\sigma-$) at $\varphi= $135$^0$ $(P_{circ}= -1)$. The generated photocurrent is measured in the $y$-direction. The phenomenological expression for the photocurrent density is given by \cite{Eginligil},
\begin{equation}
 j_{PC} = C sin2\varphi + L_1 sin4\varphi + L_2 cos4\varphi + D   
\end{equation}
The first term on the r.h.s. with coefficient C represents the contribution $j_{CPGE}$  to the photocurrent due to CPGE which utilises the spin-valley coupling and angular momentum optical selection rules to generate a valley-selective spin-coupled photocurrent. The CPGE-induced current is the dominant contributor to $j_{PC}$ for light energy on-resonance with the excitonic (mainly A) transitions and reverses its direction on reversing the polarization of the light from RCP to LCP state. The y-component of the CPGE current density is given by \cite{Ganichev},
\begin{equation}
 j_{CPGE,y} = i\sum_j \gamma_{yj} (\textbf{E} \times \textbf{E}^*)_j,  i(\textbf{E} \times \textbf{E}^*)_j = \hat{e_{j}}E^2_0 P_{circ} 
\end{equation}
where $\gamma$ is a second rank pseudo-tensor, j stands for Cartesian coordinates, $\textbf{E}$ is the complex electric field of light, $E_0$ its amplitude, $\hat{e}= \textbf{q}/q$ is the unit vector in the direction of propagation of light with $q$ being the light wavevector inside the sample and $P_{circ}= sin2\varphi$ is the helicity of the incident light varying from LCP ($P_{circ}= -1$) to RCP ($P_{circ}= +1$) illumination. The polarization-angle dependent factor of $sin2\varphi$  in the phenomenological expression for the CPGE current in Eq. (3) thus arises from $P_{circ}$. The second term on the r.h.s. of Eq. 3 with coefficient L$_1$  represents a spin-independent contribution j$_{LPGE}$  when the illuminating light is linearly polarized, termed the linear photogalvanic effect (LPGE). The origin of LPGE lies in the dissipative scattering of electrons in the sample. The third term on the r.h.s. of Eq. 3 with the coefficient L$_2$ represents the spin-independent contribution j$_{LPDE}$ due to the linear photon drag effect (LPDE) which arises from the transfer of light momentum to electrons. In this case also, the light falling on the sample is linearly polarized. The fourth term $D$ on the r.h.s. of Eq. 3 represents a polarization independent term. One notes that j$_{CPGE}$ is $\pi$ periodic whereas j$_{LPGE}$ and j$_{LPDE}$ are $\pi/2$ periodic. 
\begin{figure*}
 \centering
\includegraphics[width=0.7\linewidth]{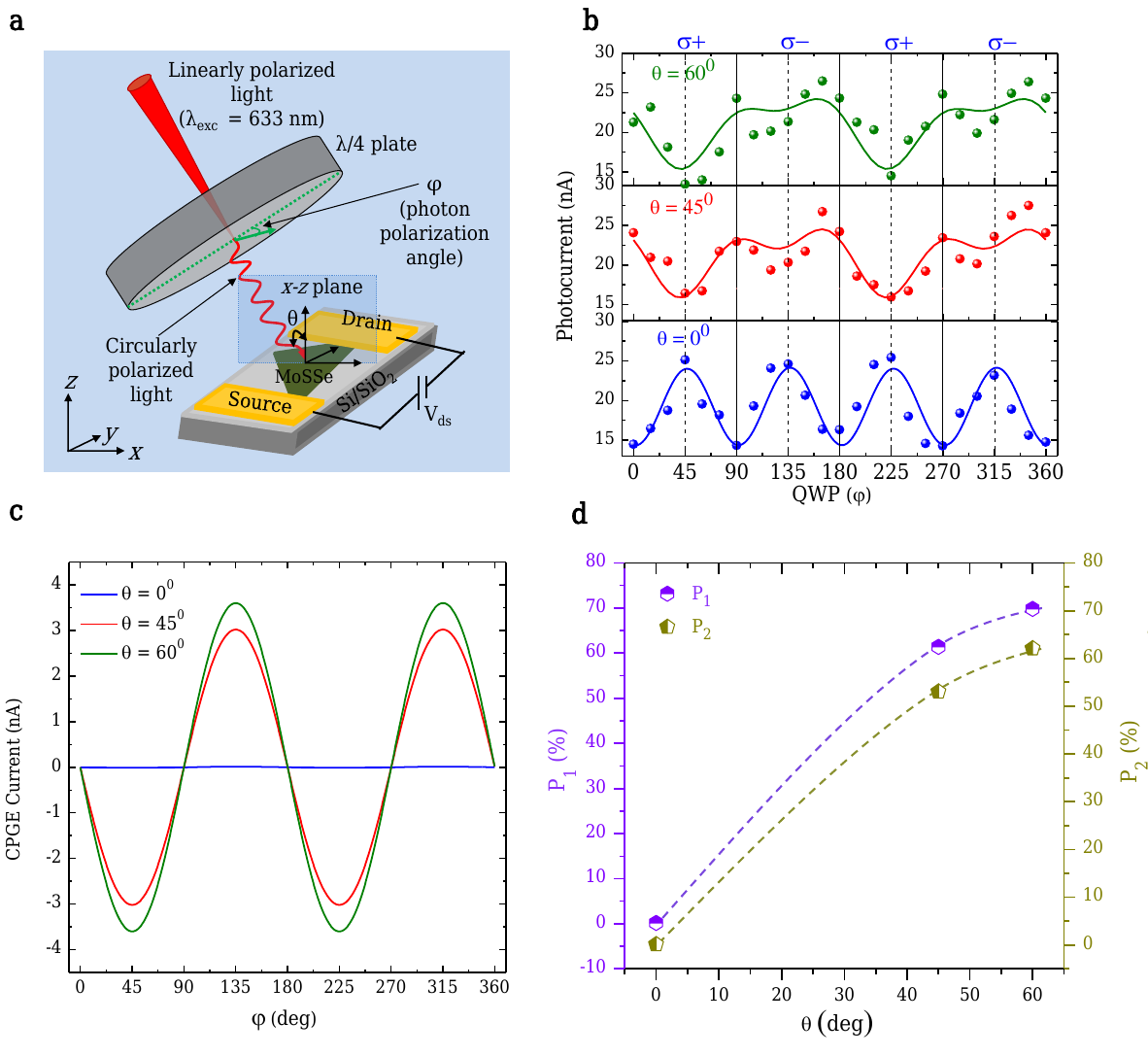}
\caption{Experimantal configuration, helicity-dependent photocurrent and degree of polarization. {\bf{a}} Schematic diagram of experimental setup in which the polarization of incoming light is changed by rotating the QWP from 0$^0$ to 360$^0$. {\bf{b}} Photocurrent versus angle of polarization $\varphi$ using near on-resonance excitation energy of 1.96 eV (angle of incidence  $\theta=$0$^0$: blue plot, 45$^0$: red plot and 60$^0$: green plot), with the solid dots representing experimental data. The solid lines are the fitting functions based on Eq. 3. {\bf{c}} Plot of the CPGE current versus $\varphi$. {\bf{d}} Degree of photocurrent polarization P$_1$ and P$_2$ for $\theta=$ 0$^0$, 45$^0$and 60$^0$ at 1.96 eV excitation energy.}
\label{Fig. 1.}
\end{figure*}
\subsubsection*{Photocurrent under near on-resonance illumination}
To explore the polarization dependent photoresponse behavior, we employed linearly polarized light with energies of 1.96 eV (near on-resonance, illumination power $=$ 75 $\mu$W) to illuminate the sample while maintaining a constant source-drain voltage V$_{ds} =$ 1V. The photocurrent which  mainly arises from the A exciton and A trion, exhibited a clear dependence on $\sigma+$ and $\sigma-$ excitation, as depicted in Fig. 3b (in case of $\theta \neq 0^0$). Three angles of incidence of the illuminating light, $\theta = $0$^0$, 45$^0$ and 60$^0$, were chosen for the photocurrent measurements. From the fitting (solid lines) of the experimental data (solid points) using Eq. 3, we obtained the values of coefficients C, L$_1$ and L$_2$. Fig. 3c shows the CPGE current $j_{CPGE} = Csin2\varphi$ versus $\varphi$ for the three angles of incidence. For the non-zero values of $\theta$, the direction of the photocurrent $j_{CPGE}$ is reversed when the polarization-angle $\varphi$ changes from 45$^0$ (RCP) to 135$^0$ (LCP) with the magnitude of the current being the largest at these $\varphi$ values. For normal incidence of light, $\theta=$0$^0$, the magnitude of the CPGE current turns out to be  zero. The maximum amplitude of the CPGE current is attained for $\theta=$ 60$^0$. Supplementary Fig. 15a shows the variation of the absolute values of the coefficients, C, L$_1$ and L$_2$ as a function of $\theta$. One can define a polarization degree of photocurrent $P$ as \cite{Ganichev},
\begin{equation}
P = \frac{I(\sigma+)-I(\sigma-)}{I(\sigma+)+I(\sigma-)}
\end{equation}
where I$(\sigma+)$ and I$(\sigma-)$ represents the photocurrent generated by RCP and LCP optical excitation respectively. Using the expression for the photocurrent in Eq. 3, the polarization degree can be negative or greater than 1 since the fitting parameters C, L$_1$ and L$_2$ can have positive or negative sign. To keep P positive as well as less than 1, one can use a new definition P$_1$ of polarization degree as,
\begin{equation}
P_1=\frac{|[Csin2(\sigma+)+ L_1sin4(\sigma+)+ L_2cos4(\sigma+)]
-[Csin2(\sigma-)+ L_1sin4(\sigma-)+L_2cos4(\sigma-)]|}{|Csin2(\sigma+)|+ |L_1sin4(\sigma+)|+|L_2cos4(\sigma+)|+ |Csin2(\sigma-)|+ |L_1sin4(\sigma-)|+|L_2cos4(\sigma-)|}
\end{equation}
where $\sigma+$, $\sigma-$ correspond to $\varphi= $45$^0$, 135$^0$ respectively. The unpolarized contribution $D$ is left out from the calculation. Plugging in the values of $\varphi$, one gets,
\begin{equation}
P_1=\frac{|C|}{|C|+|L_2|}
\end{equation}
An alternative expression used in literature \cite{Cha} is,
\begin{equation}
P_2=\frac{|C|}{|C|+|L_1|+|L_2|}
\end{equation}
This is the ratio between the absolute value of the coefficient of j$_{CPGE}$ and the sum of the absolute values of the coefficients of all the polarization-dependent contributions to the total photocurrent. The absolute values of the coefficients represent the maximum values of the corresponding components of the photocurrent. For the data reported in Fig. 3b, the values of the fitting parameters are: $C$ = -3.77nA, $L_1$= -0.67 nA, $L_2$ = 1.63 nA for $\theta =$ 60$^0$. The polarization degrees of the photocurrent are, P$_1=$ 69.8$\%$, P$_2=$ 62.1$\%$. Figure 3d illustrates the variation of P$_1$ and P$_2$ values with respect to $\theta$. These results clearly demonstrate the successful generation of a substantial photocurrent polarization, achieved naturally without the need for any external manipulation or intervention. The figure of merit of photocurrent polarization can be defined as \cite{Eginligil},
\begin{equation}
F=\frac{|C|}{|L_2|}
\end{equation}
The plot of F versus the angle of incidence of illumination light is shown in Supplementary Fig. 15b. It is observed that F reaches its maximum value for $\theta =$ 60$^0$. To check the repeatability of the helicity of light dependent photocurrent, we have performed the measurements on additional two monolayer TMDC based devices and the results are displayed in Supplementary Fig. 17.
\begin{figure*}[t!]
 \centering
\includegraphics[width=1\linewidth]{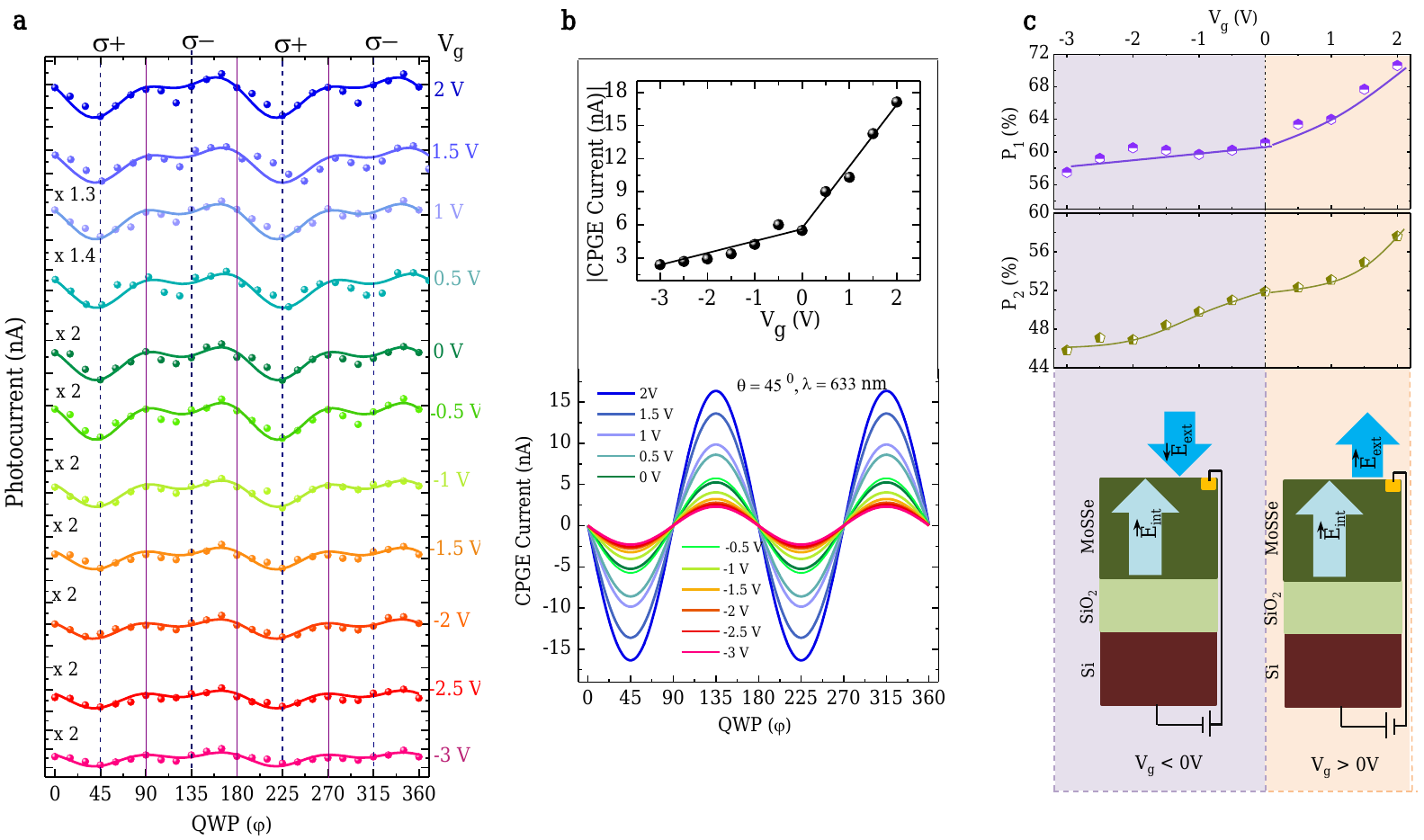}
\caption{Gate voltage-tuned helicity dependent photocurrent measurement. {\bf{a}} Photocurrent data (solid dots) with gate voltage (V$_g$) varied from -3V to +2V under near on-resonance 1.96 eV excitation with $\theta =$ 45$^0$. The solid lines are the fitting functions. {\bf{b}} Graphical representation of CPGE current ($Csin2\varphi$) as a function of QWP angle $\varphi$ for all gate voltages with $\theta =$ 45$^0$. The solid lines are the fitting functions based on the photocurrent expression in Eq. 3. Inset shows the CPGE current as a function of V$_g$ for $\theta =$ 45$^0$. {\bf{c}} Degrees of photocurrent polarization P$_1$ and P$_2$ as functions of gate voltage. Lower panel shows schematically the internal electric field E$_{int}$ and external electric field E$_{ext}$ with the latter arising from gate voltage V$_g$. }
\label{Fig. 1.}
\end{figure*}
\subsubsection*{Photocurrent under external electric field}
To explore the effects of an external perpendicular electric field on photocurrent generation under polarized light in monolayer TMDCA, an external gate voltage (V$_g$) was applied to the sample and varied from -3V to +2V at a constant drain-source voltage (V$_{ds}$) of 1V. The key results of our study are presented in Fig. 4a. We utilized a near on-resonance light source with energy of 1.96 eV (illumination power of 75 $\mu$W) and an angle of incidence ($\theta$) of 45$^0$. The specific angle of incidence is chosen to better address  experimental challenges. The CPGE current under varying gate voltage is depicted in Fig. 4b. Figure 4c demonstrates the variation of the photocurrent polarization degree P$_1$ and P$_2$ with gate voltage. The absolute values of fitted parameters (L$_1$ and L$_2$) and F value as a function of gate voltage are displayed in Supplementary Fig. 16a and b, respectively. This study demonstrates the tunability of the photocurrent, the CPGE current and the photocurrent polarization degree, by varying the external voltage. As V$_g$ increases beyond 0V, both the CPGE current and the P$_1$ or P$_2$ values are enhanced significantly. Conversely, for V$_g<$ 0V, both the values decrease gradually. The experimental data indicate that a positive gate voltage enhances the effect of the internal field $\mathrm{E_{int}}$ in the generation of a photocurrent as both are aligned in the positive $z$-direction, shown schematically in the lower panel of Fig. 4c. The oppositely-directed negative gate voltage, on the other hand, reduces the effect of the internal field. The computed dipole moment (per unit cell) via first principles calculations as a function of the external electric field, $\mathrm{E_{ext}}$, is shown in Supplementary Fig. 4b. We explored the possibility of the contribution of the Rashba effect to the generation of the photocurrent in the TMDCA material. The generation of a CPGE current due to direct interband transitions at the $\Gamma$ point has been reported in experiments involving quantum well structures \cite{Ganichev}. The electronic energy band structures of both monolayer and bilayer TMDCA (Fig. 2a and Supplementary Fig. 6b) rule out the possibility of direct transitions as the lowest conduction band at the $\Gamma$ point lies much higher than the CB minimum at the K points. The other possibility lies in an indirect band-gap transition from the VB branches at the $\Gamma$ point to the CB at the K point \cite{Patel, Du}. PL measurements (Supplementary Fig. 18) for excitation energy of 2.54 eV  rule out this possibility in the case of a few-layer TMDCA material as the PL intensity spectrum shows that direct transitions are significantly more favorable than  indirect ones. In the limit of the bulk material, the direct and the indirect transitions are both favoured.

\subsubsection*{Photocurrent under off-resonance illumination}
In a subsequent experiment, we replicated the procedure of obtaining polarization-resolved PL spectra using a near off-resonance excitation of 2.54 eV. The PL spectra are depicted in Supplementary Fig. 19, exhibiting a valley polarization degree of $\eta=$ 30.6$\%$ (Eq. 2). The result is in sharp contrast with that obtained in the case of monolayer MoS$_2$ using a polarized excitation of 2.33 eV, off-resonance with the A or B exciton \cite{Shan}. The observed emission was found to be fully unpolarized with $\eta=$ 0$\%$. The degree of polarization of the PL was found to be high when the excitation was on-resonance but had a power-law decrease as the excitation energy increased \cite{Kioseoglu}. Phonon-assisted intervalley scattering has been identified as the primary mechanism for valley depolarization. A finite value of $\eta$ for the monolayer TMDCA when the excitation is off-resonance signifies that the valley polarization is more robust in the case of the alloy material. A similar feature was observed in the case of the monolayer TMDC material WS$_2$ using an off-resonance polarized excitation of 2.33 eV, with $\eta=$ 16$\%$ at a temperature of 10K \cite{Zhu}. The value of  $\eta=$ 30.6$\%$ obtained in the case of the alloy material at room temperature signifies a comparatively more robust valley polarization. Unlike in the case of monolayer MoS$_2$, the depolarization mechanism may not be solely dictated by the off-resonance condition. Supplementary Fig. 20a shows the plot of photocurrent versus the photon polarization-angle $\varphi$ with the angles of incidence $\theta =$ 60$^0$, 45$^0$, 0$^0$ using an off-resonance  excitation of 2.54 eV. For the first two angles of incidence, there is significant polarization between the $\sigma+$ and $\sigma-$ excitations with the degrees of photocurrent polarization given by P$_1 =$ 52.5$\%$ and P$_2 =$ 36.7$\%$. For near on-resonance excitation of 1.96 eV, P$_1 =$ 69.8$\%$ and P$_2 =$ 62.1$\%$. In comparison, the degree of photocurrent polarization is negligible (less than 1$\%$) in the case of monolayer MoS$_2$ when the excitation is off-resonance \cite{Eginligil}. Supplementary Fig. 20b shows the CPGE current, j$_{CPGE}$, versus $\varphi$ for three angles of incidence in the off-resonance condition. The magnitude of the CPGE current, for $\theta =$ 60$^0$ and 45$^0$, is non-negligible, in contrast to the case of monolayer MoS$_2$. It is important to note that the $\eta$ values obtained from PL measurements, for both near on-resonance and off-resonance  excitations, are smaller compared to the polarization degree of the photocurrent. This can be attributed to the loss of polarization during excitation with circularly polarized light and the subsequent collection in PL measurements \cite{Eginligil}. Room-temperature valley polarization under off-resonance excitation in TMDCA offers a new paradigm for valleytronics applications. 

\subsubsection*{Photocurrent in bilayer TMDCA}
To provide a comparative analysis, we conducted identical measurements on a bilayer TMDCA-based device (the height profile is illustrated in Supplementary Fig. 21a) without the application of any external electric field. The system was subjected to circularly polarized excitation with an energy of 1.96 eV. The polarization-dependent photocurrent measurement and the values of the fitted parameters are summarized in Supplementary Fig. 21b and c respectively. The degrees of photocurrent polarization, P$_1$, P$_2$, and the F are calculated to be 70.4$\%$, 55$\%$ and 2.38 respectively with the angle of incidence $\theta =$ 45$^0$. To gain a deeper understanding of the underlying mechanisms contributing to the results, we performed DFT calculations for the bilayer TMDCA (Se = 25$\%$, S = 75$\%$). Unlike its parent TMDCs such as MoS$_2$ and MoSe$_2$, the asymmetric arrangement of S and Se atoms on both sides of the Mo atoms in the bilayer alloy confers on it the character of a non-centrosymmetric crystal. The DFT calculations elucidate the spin-valley physics of the bilayer, akin to the monolayer, revealing a noteworthy splitting of the valence band (VB) with in-plane spin-orientations of the band electrons and Rashba splitting around the $\Gamma$ point, as illustrated in Supplementary Fig. 6b-e. The generation of valley polarization in the bilayer material, even in the absence of an external electric field, considerably enhances its application potential. 
\begin{table*}[t]
\centering
\caption{\bf{Comparison of helicity-dependent photocurrent measurements }}
\includegraphics[width=1\linewidth]{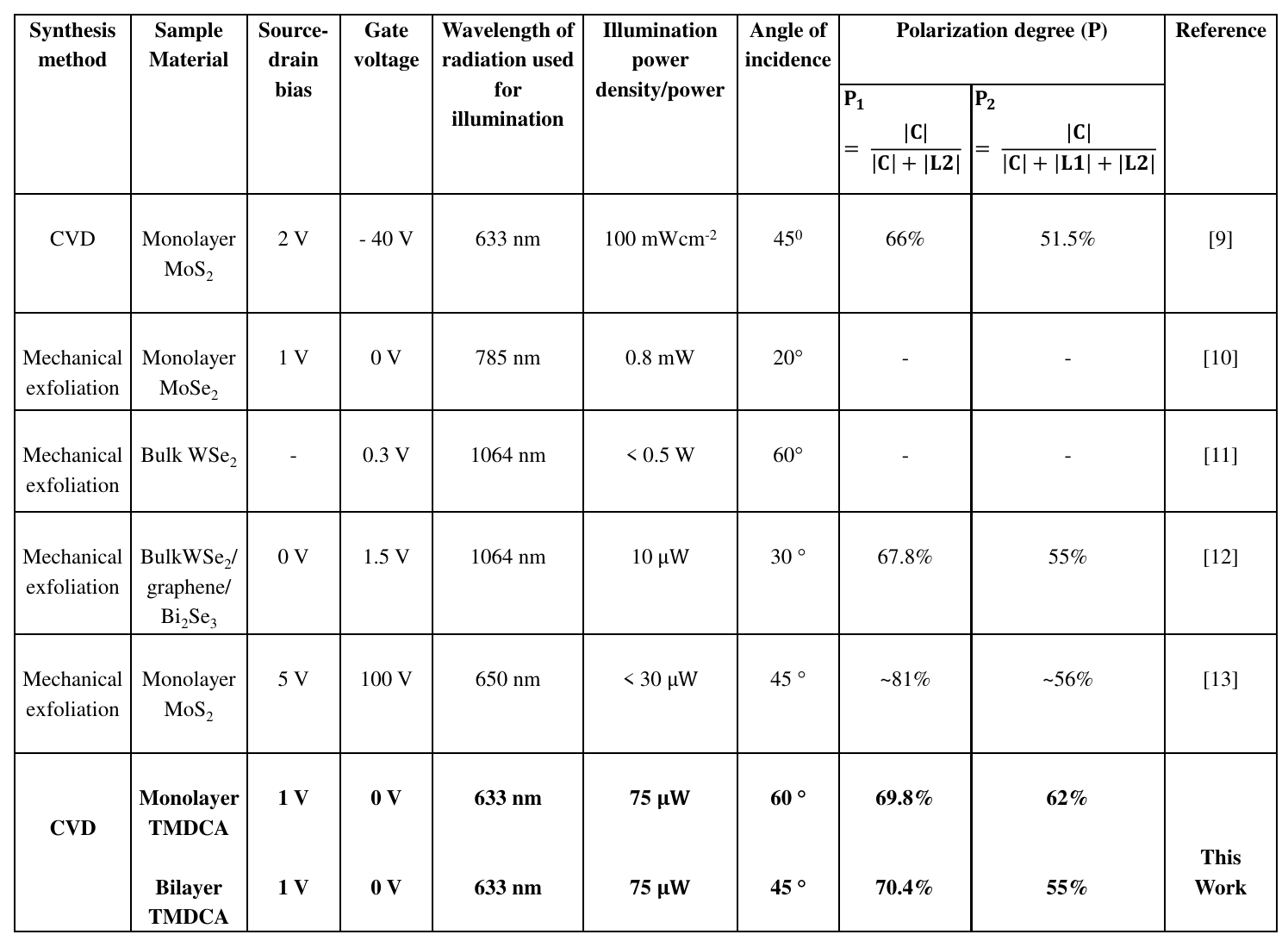}
\end{table*}

\subsection*{Discussion}
Our study, involving both experiments and first principles calculations, establishes that the TMDCA monolayer material, MoS$_{2x}$Se$_{2(1-x)}$ (x=0.6), exhibits the characteristic features of spin-valley physics at the K and K$^{\prime}$ points and a Rashba splitting of the VB at the $\Gamma$ point. We have provided a comparison in Table 1 of the performance of our  device vis-à-vis other similar devices reported in literature. As shown in the table, the range of gate voltages required to modulate the photocurrent in the TMDCA material is small, about a few volts. With a comparatively low laser power density, our photocurrent set-up could generate  photocurrents  in the nA range with large degrees of photocurrent polarization at room temperature. The significant advantages, already pointed out, call for a wider exploration of TMDCA and related Janus materials as possible device materials in valleytronics applications. 

A fundamental quantity which differentiates between the K  and K$^{\prime}$ valleys is the Berry curvature (BC) $\bf{\Omega}$, interpreted as an effective magnetic field in momentum space \cite{Schaibley, Xiaodong, Ziao}. The BC is a pseudovector with an out-of-plane orientation and opposite signs at the two valleys. The valley Hall effect (VHE), demonstrated experimentally in monolayer  MoS$_2$ \cite{k}, has its origin in the opposite signs of the BC at the two valleys so that the electrons belonging to these valleys move in opposite transverse directions in the presence of an in-plane electric field. An externally applied magnetic field is not needed to observe this Hall-like phenomenon. In the case of gated/polar TMDCS, the electrons at the K, K$^{\prime}$ valleys experience both Zeeman-type (Ising) and Rashba SOIs. For such materials, an intriguing proposal has been made of a novel type of VHE termed spin-orbit coupling induced spin Hall effect (SVHE) \cite{Zhou} in which a significant enhancement of the spin-type BC over the orbital BC at the CB edge, under tunable external gating, could result in a prominent SVHE. A recent theoretical study \cite{Yu} has shown that the the Rashba SOI enhances the spin Hall effect in the monolayer Janus material MoSSe. The experimental demonstration \cite{Moore, Queredaj} of superconductivity, with a highest T$_c$  of 10K, in gated thin films of MoS$_2$ is another phenomenon the origin of which lies in the coexistence of Ising and Rashba SOIs, an intrinsic feature of the Janus and TMDCA materials. The possibility of the contribution of the BC to a CPGE photocurrent was first discussed in a theoretical study \cite{Moore}. In this case, the CPGE current j$_{CPGE}$ has a $\theta$-depenence of $cos\theta$ so that the BC contribution is maximized for  $\theta=$ 0$^0$, i.e., for normal incidence of the illuminating light \cite{Quereda}. Experiments involving monolayer TMDCs as well as TMDCA rule out a BC contribution to the CPGE current as the current has finite values only for oblique angles of light incidence, i.e., $\theta \neq$ 0$^0$. A recent study \cite{Queredaj} provides examples of experimental conditions which allow for the generation of a CPGE current at normal incidence of light. Both the monolayer TMDCA and the monolayer Janus MoSSe possess an intrinsic electric field perpendicular to the monolayer surface. The field contributes to the generation of the Rashba SOI, at the core of applications in spintronics and optoelectronics. The added advantage of the TMDCA is its ease of synthesis making it more suitable for device applications. The TMDCA materials have so far received scant attention from researchers. It is to be hoped that our experimental and computational results on a specific TMDCA material would stimulate further research on polar TMDC materials like Janus and TMDC alloys.

\subsection*{Methods}
\subsubsection*{Sample growth}
TMDCA monolayer and bilayer samples were synthesized on a SiO$_2$/doped Si substrate using the chemical vapor deposition (CVD) method in a homemade horizontal tube furnace connected to a vacuum pump. A schematic representation of the setup is presented in Supplementary Fig. 7a. During the synthesis process, we optimized various parameters, including the distances between the Mo, S, and Se sources, temperature, and flow rates. A quartz boat containing Mo powder was placed within the tube in a high-temperature zone ($\sim$ 750$^0$C), while the S and Se powders were positioned in a relatively low-temperature zone ($\sim$ 200$^0$C). The temperature of the furnace was gradually increased at the rate of 25$^0$C/min with a flow 10 sccm of high-purity Ar gas, reaching its maximum temperature over a span of 30 minutes. The furnace was maintained at the maximum temperature for 10 minutes with 40 sccm Ar flow before being allowed to cool down naturally to room temperature.
\subsubsection*{Characterization Tools}
The morphology of the as-grown samples was analyzed using a FEI Quanta 200 SEM. The SEM was equipped with an EDX spectrometer for compositional analysis. To determine the layer number and topography, AFM was performed using a Bruker instrument (Innova). PFM measurements were conducted using an atomic force microscope equipped with a dual AC resonance tracking piezoresponse module (MFP-3D Origin, Asylum Research) in contact mode. The vibrational and optical properties of the samples were investigated through Raman and PL measurements. For the experiments, a micro-Raman spectrometer (LabRAM HR, Horiba Jobin Yvon) coupled with a Peltier-cooled CCD detector was utilized. An air-cooled argon-ion laser (Ar+) with a wavelength of 488 nm and a 633 nm He-Ne laser were employed as excitation light sources. The laser beam was focused on the sample using a 50x objective with a numerical aperture (NA) of 0.75.
\subsubsection*{Device Fabrication}
To prepare the electrical contacts on the sample, freshly prepared CVD grown TMDCA samples on a highly doped Si/SiO$_2$ substrate was initially cleaned using a standard method involving acetone and IPA cleaning. The cleaned substrates were then dried using nitrogen blowing and subjected to a brief heating on a hotplate to remove any remaining moisture. Next, the samples were spin-coated with AZ1512-HS positive photoresist and baked at 100$^0$C for one minute. Subsequently, the appropriate flakes were selected, and a laser writer (Microtech, Model LW405) with a laser power of 160 mJ/cm$^2$ was utilized to design the pattern for the electrical contacts. Following the patterning step, a deposition of Cr/Au (5 nm/60 nm) was carried out using an e-beam evaporation system. The resist was then lifted off using acetone to reveal the desired electrical contact pattern. The resulting fabricated devices were subsequently bonded using gold wire and silver epoxy paste. Throughout the entire device fabrication process, strict measures were taken to maintain proper translucence and minimize the risk of contamination. 
\subsubsection*{Optoelectronic measurements}
Photocurrent measurements were conducted using a two-probe setup with a Keithley 2450 SMU while subject to the illumination of two coherent light beams with energies of 1.96 eV and 2.54 eV. For helicity-dependent photocurrent measurements, a half-wave plate ($\lambda/2$ plate) followed by a QWP ($\lambda/4$ plate) were employed to convert the linearly polarized light beams into CPL. A 10x objective lens was utilized to focus the light onto the sample surface. Neutral density filters were applied to adjust the illumination intensity. The measurements were carried out at lower power densities to mitigate any potential light-induced heating effects. To generate spin-valley-selective circular current, the QWP was systematically rotated from 0$^0$ to 360$^0$. All measurements were conducted at room temperature and under ambient condition. 

\newpage
\renewcommand{\figurename}{Supplementary Figure}
\setcounter{figure}{0}
\setcounter{table}{0}
\renewcommand\tablename{Supplementary Table}
\setcounter{equation}{0}
\section*{Supplementary Information} 
\subsection*{Supplementary Note 1: Theoretical derivation and first principles calculations}
\subsubsection*{I. Model Hamiltonians illustrating spin-valley coupling and Rashba physics}	 

The minimal k.p Hamiltonian close to the K  and K$^{\prime}$ points is given by \cite{Patel, Gmitra},
\begin{equation}
H(k) = at(\tau k_x\sigma_{x} + k_y \sigma_{y})+ \Delta \sigma_{z} + \lambda_{V} \tau \frac{1-\sigma_{z}}{2}S_z + \lambda_{C}\tau\frac{1+\sigma_{z}}{2}S_z
\end{equation}
where $a$ is the lattice constant, $t$ the hopping integral, $\tau$ is the valley index with value +1 (-1) at the K (K$^{\prime}$) valley, 2$\lambda_{V}$ (2$\lambda_{C}$ ) are the magnitudes of Zeeman-type spin splittings of the uppermost (lowermost) VB (CB), $k_x$, $k_y$  are the Cartesian components of the electron momentum with $\textbf{k}$ measured from the K, K$^{\prime}$ points, $\Delta$ represents the direct band-gap, $\sigma_{x}$, $\sigma_{y}$ are the pseudospin Pauli matrices acting on the subspace of $d$ orbitals and $S_z$ is the z-component of the electron spin operator. Due to the decoupling of the spins, the eigenvalues of the spin operator are good quantum numbers. Considering, $\tau =$ 1, the first term in the Hamiltonian disappears at the K point and the Hamiltonian is in a diagonal form with the eigenvalues, in order of decreasing energy, given by
\begin{equation}
E_{CB,\uparrow} = \Delta + \lambda_{C}, E_{CB,\downarrow} = \Delta - \lambda_{C}, E_{VB,\downarrow} = -\Delta + \lambda_{V}, E_{VB,\uparrow} = -\Delta - \lambda_{V} 
\end{equation}
The sequence of the split subbands with the associated out-of-plane electron spin-orientations are as depicted in Fig. 1a and b and is supported by first principles DFT calculations (Fig. 2). For the K$^{\prime}$ valley with $\tau =$ -1, the energy eigenvalues are the same but the up-and down-spin-orientations are interchanged because of TRS. Each valley has its own unique sequence of electron spin-orientations indicating that the valley and spin degrees of freedom are coupled. The coupling extends over the whole of the K and K$^{\prime}$ valleys as shown by our first principles calculations.

An electron moving in a plane and subjected to a vertical electric field $\textbf{E}$ experiences an effective in-plane magnetic field which couples to the electron spin-components in the plane, giving rise to the Rashba SOI \cite{Patel, Bentmann}. The interaction Hamiltonian is given by,
\begin{equation}
H_R = \alpha_R(k_y\sigma_x - k_x\sigma_y)  
\end{equation}
where $\alpha_R$, the strength of the Rashba interaction, is a function of the electric field, the momentum wave vector $\textbf{k}=(k_x, k_y,0)=k(cos\vartheta,sin\vartheta,0)$ and $\sigma_x$, $\sigma_y$ are the Pauli spin matrices. In matrix form, 
\begin{equation}
H_R = \alpha_R
\begin{pmatrix}
0 & k_y+ik_x\\
k_y-ik_x & 0
\end{pmatrix}
\end{equation}
In the absence of the Rashba interaction, the parabolic energy band at the high symmetry $\Gamma$ point, with energy dispersion $\frac{{\hbar^2}{k^2}}{2m^*}$ ( $m^*$ is effective mass), is spin-degenerate. The Rashba SOI removes the spin-degeneracy by splitting the energy band into two branches, with a linear shift in momentum-space from the original band. The eigenvalues of $H_R$ are $\lambda_{R1}= +\alpha_{R}k$, $\lambda_{R2}= -\alpha_{R}k$, $k=\sqrt{{k_x}^2+{k_y}^2}$ with $k=$0  at the $\Gamma$ point. These are the Rashba SOI-split energy eigenvalues of  $H_R$ with  the corresponding eigenvectors given by $u_{R1} = \frac{1}{\sqrt{2}}\begin{pmatrix}
1\\ -ie^{i\vartheta}\end{pmatrix}$, $u_{R2} = \frac{1}{\sqrt{2}}\begin{pmatrix}
-ie^{-i\vartheta}\\ 1\end{pmatrix}$. The  total energy eigenvalues are  $E_{tot} = \frac{{\hbar^2}{k^2}}{2m^*}{\pm}\alpha_{R} k$. Fig. 2 provides clear evidence of the Rashba spin splitting of the topmost VB of the monolayer TMDCA at the $\Gamma$ point. The spin polarizations are given by the expectation values:
\begin{equation}
{<\bm{\sigma}>}_{u_{R1}} =
\begin{pmatrix}
sin\vartheta, & -cos\vartheta, & 0\\
\end{pmatrix},
{<\bm{\sigma}>}_{u_{R2}} = 
-\begin{pmatrix}
sin\vartheta, & -cos\vartheta, & 0\\
\end{pmatrix},
\end{equation}
We note that ${<\bm{\sigma}>}_{u_{R1}} = -{<\bm{\sigma}>}_{u_{R2}}$. The expectation values indicate that the spins are aligned in-plane with orientations perpendicular to the wave vector $\textbf{k}= \begin{pmatrix} cos\vartheta, & sin\vartheta, & 0\\ \end{pmatrix}$ (locking of the spin and momentum directions). The in-plane spin textures around the $\Gamma$ point are confirmed by the first principles calculations for the TMDCA materials ( supplementay Fig. 4a).

\subsubsection*{II.	Computational Details of the first principles calculations}
First-principles calculations are performed using density functional theory or DFT as implemented in the Quantum ESPRESSO package with the plane wave basis set and pseudo-potentials \cite{Giannozzi}. To treat interaction between the electrons, generalized gradient approximation (GGA) has been used as the exchange-correlation functional \cite{Perdew}. The coupling between electron spin and its orbital angular momentum is considered by including the fully relativistic effect in this study. The electron-ion interactions are described by the fully-relativistic projector augmented wave pseudo-potentials \cite{Projector}. The plane-wave cut-offs for the kinetic energy and charge density for all our calculations are 50 Ry and 500 Ry, respectively. We use the $\Gamma$-centered Monkhorst-Pack k-point grid of 24 $\times$ 24 $\times$ 1. To model MoSSe alloy (TMDCA)  with Se 38$\%$ and S 62$\%$, a 2 $\times$ 2 $\times$ 1 cell of  monolayer TMDCA is constructed. The optimized equilibrium in-plane lattice parameter for TMDCA is 6.42 \AA. During structural relaxations the atomic coordinates are relaxed until the force on each atom becomes smaller than 0.1 mRy/Bohr. In order to obtain the spin texture of a given energy eigen state over the complete Brillouin zone we modify some post processing tools of Quantum ESPRESSO, as explained in \cite{Ghosh}. The external electric field is modeled using a saw-like potential alon
g the out-of-plane or z direction of the TMDCA. Dipole correction is applied to avoid spurious interactions between the periodic images of TMDCA along the direction of the external electric field \cite{Binggeli}.The spin and orbital-projected band structure and the corresponding weights are obtained from the atomic pseudo-orbital projections of the states for the atoms in the layer.

\subsubsection*{III. Conduction band splitting in monolayer TMDCA}
\begin{figure*} [h!]
 \centering
\includegraphics[width=0.55\linewidth]{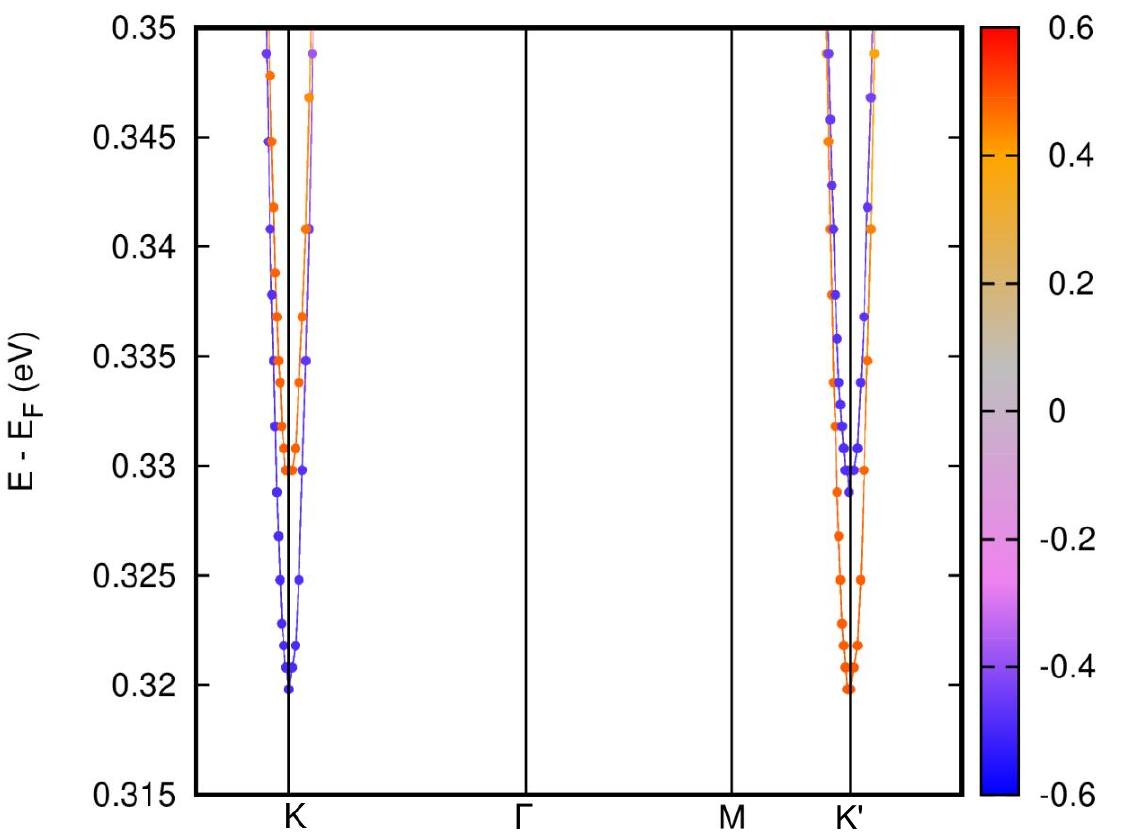}
\caption{ Zoomed view of conduction band at K and K$^{\prime}$ points which shows a slight splitting in the order of 10 meV.}
\label{Supplementary Figure 1.}
\end{figure*}
\newpage
\subsubsection*{IV.	Electronic structure of TMDCA projected on in-plane spins}
Supplementary Fig. 2a and b show the fully-relativistic band structure for TMDCA projected on $<s_x>$ and $<s_y>$. The right panel shows magnified view of the splitting between the first and second valence bands close to $\Gamma$. The in-plane spin-components have significant contribution close to $\Gamma$ where the Rashba splitting occurs.
\begin{figure*} [h!]
 \centering
\includegraphics[width=0.6\linewidth]{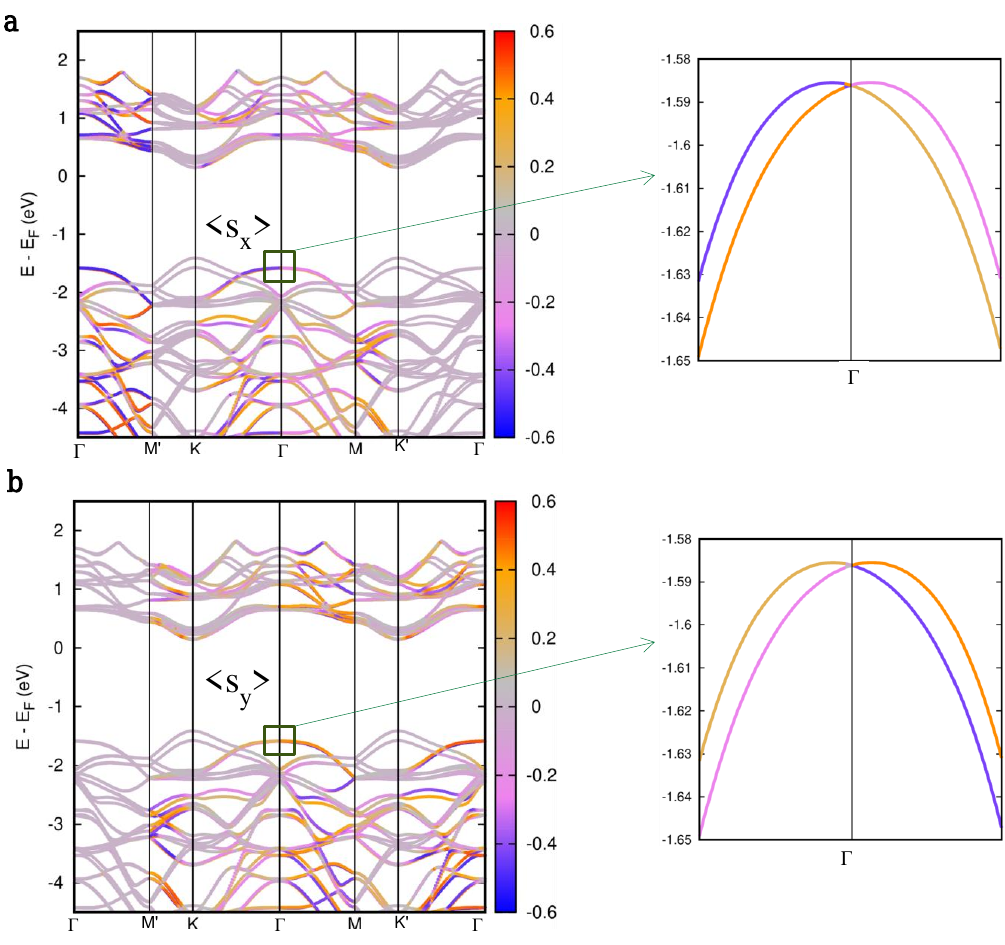}
\caption{ {\bf{a}} and {\bf{b}} show the fully-relativistic band structure plots for TMDCA projected on $<s_x>$ and $<s_y>$, respectively. The right panel shows magnified views for Rashba splitting close to $\Gamma$. The color palette shows the magnitude of in-plane spin-components along high-symmetry directions.}
\label{Supplementary Figure 1.}
\end{figure*}
\subsubsection*{V. Electronic band structure of TMDCA projected on d orbitals}
\begin{figure*} [h!]
 \centering
\includegraphics[width=0.75\linewidth]{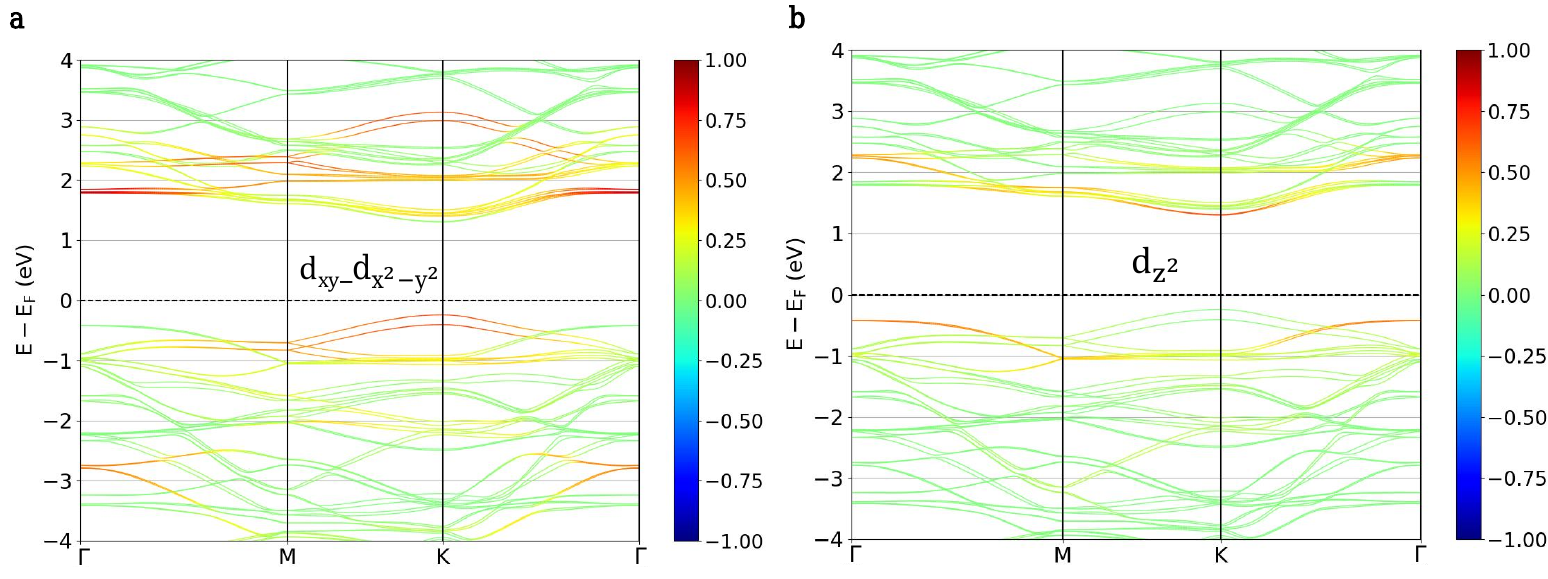}
\caption{ {\bf{a}} and {\bf{b}} show the fully-relativistic band structure plots for TMDCA projected on $d_{xy}-d_{x^2-y^2}$ and $d_{z^2}$ orbitals, respectively.}
\label{Supplementary Figure 1.}
\end{figure*}

\newpage
\subsubsection*{VI. Rashba in TMDCA}
The Rashba SOI results in the canting of in-plane spin-components in momentum-space \cite{Bihlmayer, Chen} (see Supplementary Notes 1 (I)). The spin texture of the topmost VB in the whole Brilluoin zone of the TMDCA is plotted in the left panel of Supplementary Fig. 4a with the red (blue) colour associated with the up-spin (down-spin) orientation of the electron spin at the K and K$^{\prime}$ valleys. In the central region around the $\Gamma$ point, the vectors representing the direction and magnitude of the spins are in-plane with negligible contributions from the s$_z$ component of the spins and forming a counter-clockwise rotation pattern around the $\Gamma$ point.
\begin{figure*} [h!]
 \centering
\includegraphics[width=1\linewidth]{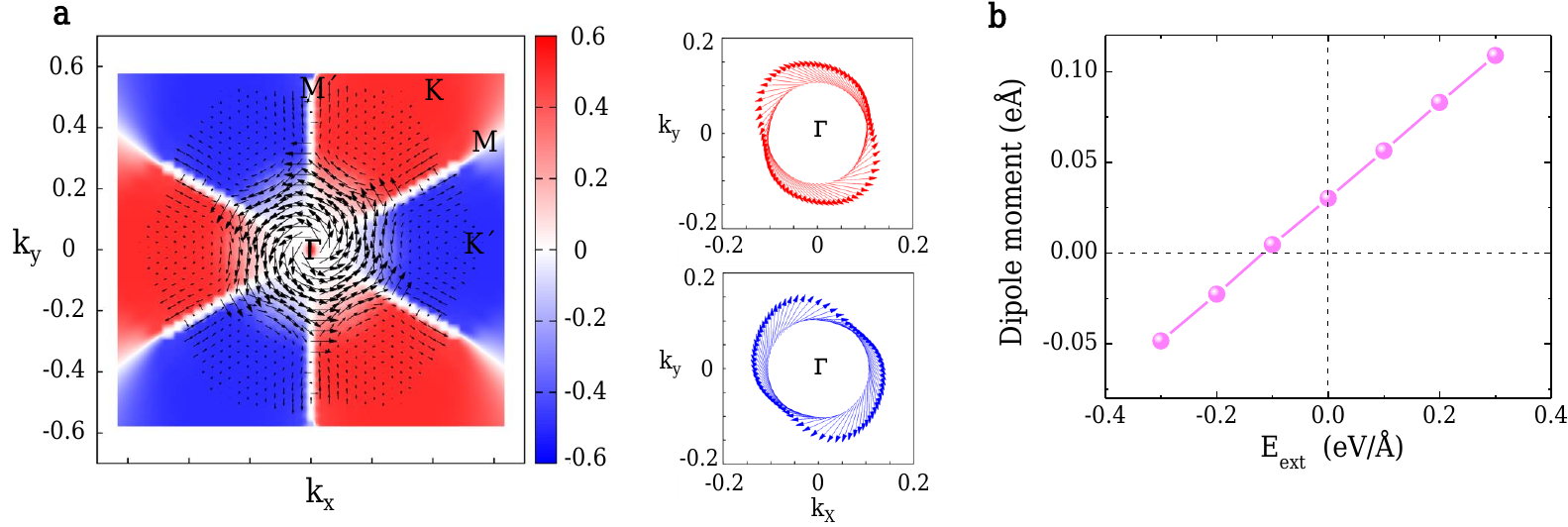}
\caption{{\bf{a}} Left panel shows 2D mapping of the in-plane and out-of-plane spin-components in the Brillouin zone for the first valence sub-band. Right panel shows the in-plane spin textures for the first (higher-energy split branch ) and second valence sub-bands over a constant energy arc with $\Gamma$ at the center. {\bf{b}} Plot of dipole moment (per unit cell) as a function of external electric field.}
\label{Supplementary Figure 1.}
\end{figure*}
\begin{figure*} [h!]
 \centering
\includegraphics[width=0.7\linewidth]{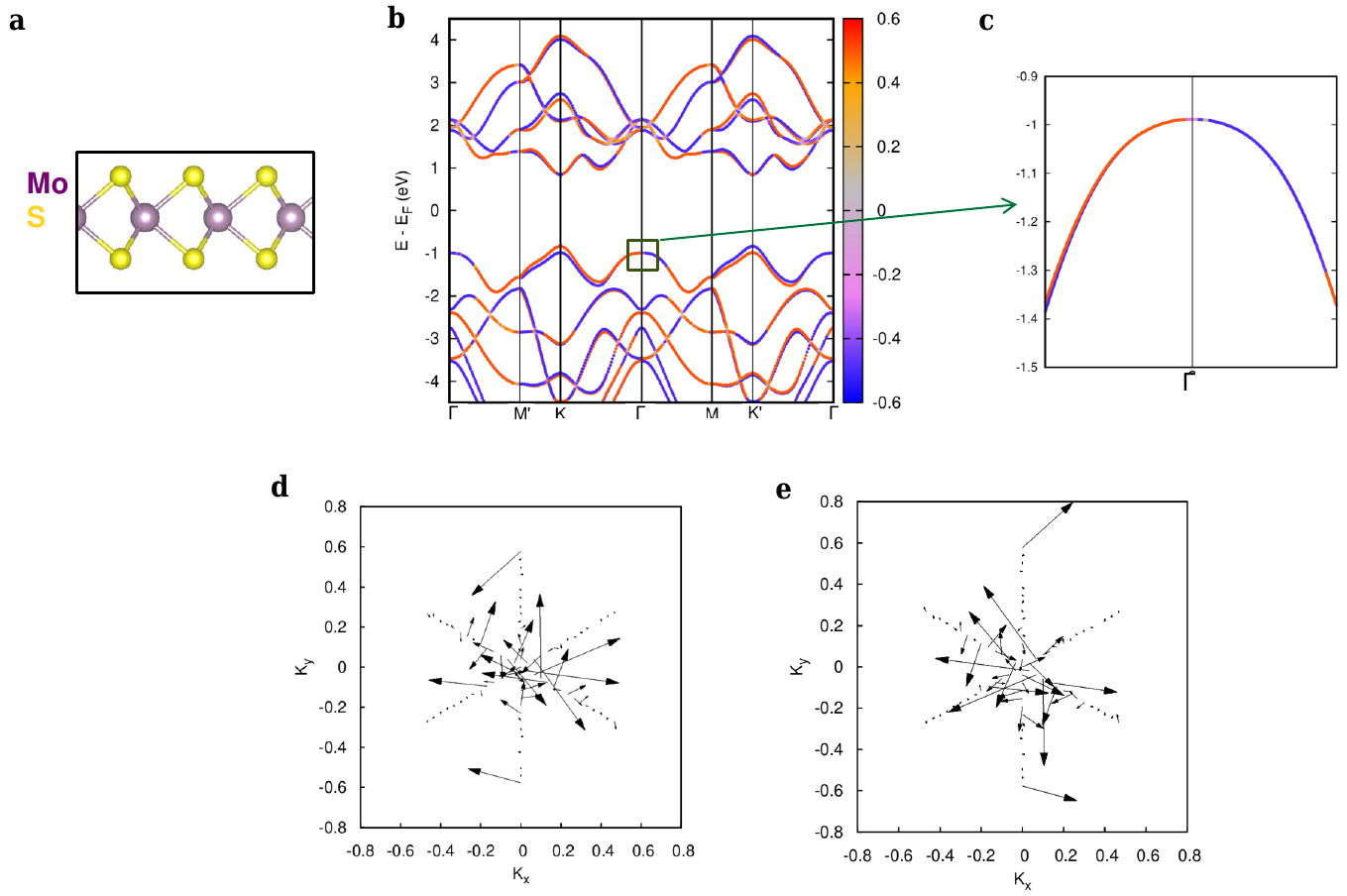}
\caption{ {\bf{a}} Side view of MoS$_2$ monolayer. {\bf{b}} Electronic band structure plot with spin-orbit coupling projected on $<s_z>$. {\bf{c}} The magnified view close to $\Gamma$, showing non-Rashba effect. In-plane spin texture plots for the {\bf{d}} first and {\bf{e}} second valence sub-bands.}
\label{Supplementary Figure 1.}
\end{figure*}

\subsubsection*{VII. Electronic band structure and spin texture of monolayer MoS$_2$}
Supplementary Fig. 5a shows the atomistic structure of MoS$_2$ monolayer, the yellow and purple atoms show S and Mo, respectively. It also depicts the fully-relativistic electronic band structure with spin-orbit coupling. Supplementary Fig. 5b depicts electronic band structure plot with spin-orbit coupling projected on $<s_z>$ while c shows the magnified version of the first and second valence bands, which are degenerate at $\Gamma$ and nearly-degenerate close to it. In-plane spin texture plots for the Supplementary Fig. 5d first and e second valence sub-bands. The MoS$_2$ monolayer shows valley splitting and valley polarization between first two valence bands. In the case of MoS$_2$ monolayer the amount of valley splitting for valence and conduction bands near the band edge is similar as obtained for the Janus MoSSe monolayer and TMDCA. The MoS$_2$ monolayer does not exhibit any internal electric field, therefore does not show Rashba splitting close to $\Gamma$. Unlike TMDCA, the energy eigen states close to $\Gamma$ are spin-polarized.

\subsubsection*{VIII. Electronic band structure and spin texture of bilayer TMDCA.}
Supplementary Fig. 6a shows the side view of bilayer TMDCA with Se$\colon$25$\%$ and S$\colon$75$\%$. The fully relativistic electronic band structure in the presence of spin-orbit coupling is plotted in Supplementary Fig. 6b, the color palette shows $<s_z>$ projection. The valley splitting between first and second valence bands is 0.16 eV, similar to the monolayer. In Supplementary Fig. 6c the Rashba splitting between first two valence sub-bands is magnified. The in-plane and out-of-plane spin textures for the first valence band are plotted in Supplementary Fig. 6d and e, respectively. Similar to the monolayer TMDCA, in this case of also the in-plane textures show Rashba effect with stronger spin moments close to $\Gamma$.

\begin{figure*} [h!]
 \centering
\includegraphics[width=0.9\linewidth]{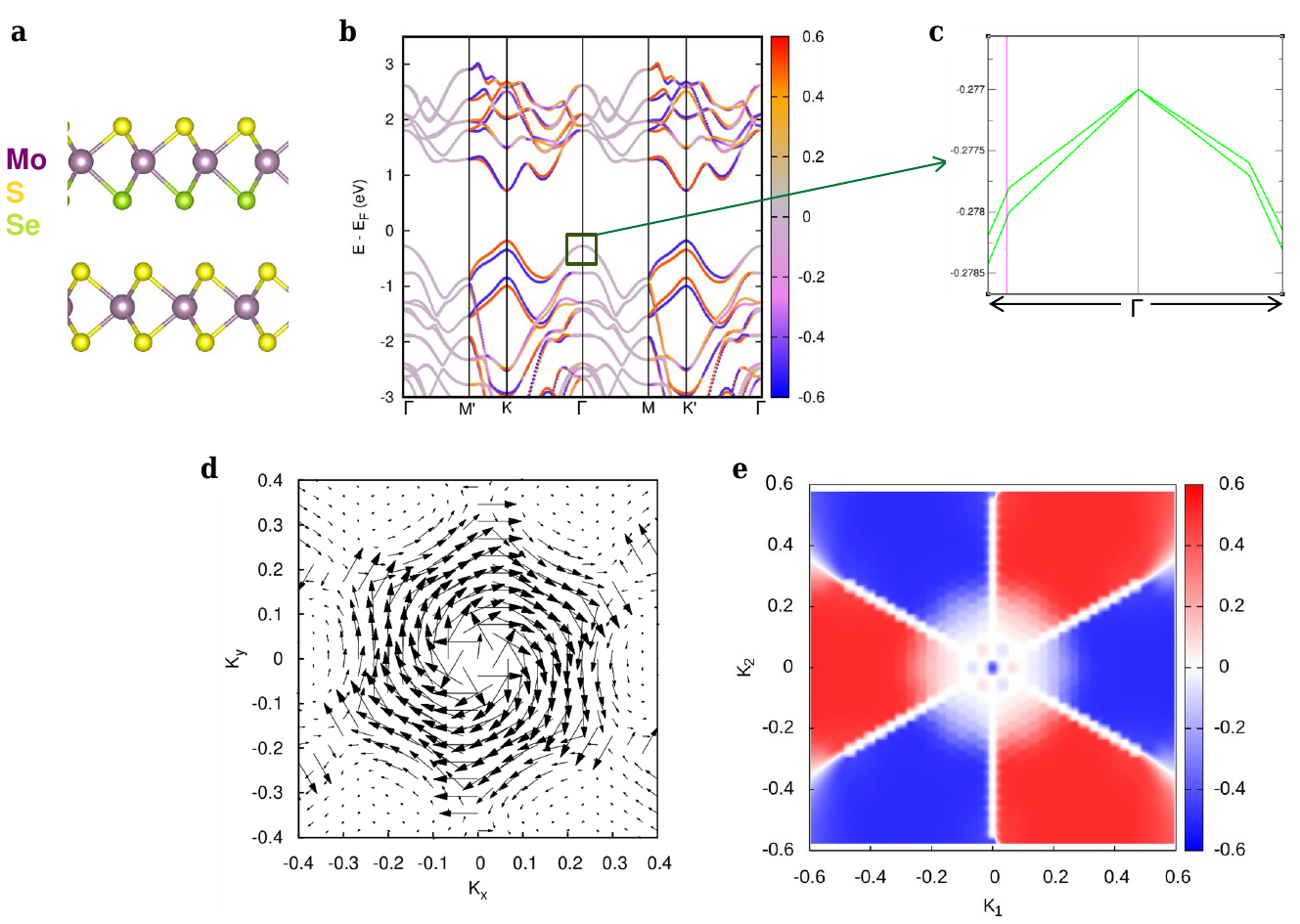}
\caption{{\bf{a}} 
 Side view of the bilayer TMDCA. {\bf{b}} Fully-relativistic band structure plot projected on $<s_z>$. {\bf{c}} Shows the magnified view of Rashba-splitting close to $\Gamma$. The {\bf{d}} in-plane and {\bf{e}} out-of-plane spin texture plots for bilayer TMDCA in the momentum-space.}
\label{Supplementary Figure 1.}
\end{figure*}

\newpage
\begingroup
\begin{table*}[h!]
{
\caption{The direct band-gap, indirect gap between $\Gamma$ and K and valley splitting between the first and second valence states.}
\centering
\begin{tabular}{|c|c|c|c|}
\hline
$E_\mathrm{field}$ & Direct gap (eV)  & $\Delta E_{\Gamma-K}$ (eV)  & VB$_\mathrm{splitting}$ (eV) \\ 
\hline
-0.4 & 1.57 & 1.70 & 0.17   \\
\hline 
-0.3 & 1.56 & 1.70 & 0.16   \\
\hline
-0.2 & 1.55 & 1.72 & 0.17   \\
\hline
-0.1 & 1.56 & 1.76 & 0.16   \\
\hline
-0.05 & 1.56 & 1.77 & 0.15  \\
\hline
0 & 1.56 & 1.78 & 0.15   \\
\hline
0.05 & 1.56 & 1.77 & 0.16 \\
\hline
0.1 & 1.56 & 1.75 & 0.16  \\
\hline
0.2 & 1.56 & 1.70 & 0.16 \\
\hline
0.3 & 1.56 & 1.68 & 0.16  \\
\hline
0.4 & 1.50 & 1.65 & 0.15   \\
\hline

\end{tabular}
\label{Energy}
}
\end{table*}
\endgroup

\newpage
\subsection*{Supplementary Note 2: CVD synthesis and primary characterizations of monolayer TMDCA}

\begin{figure*} [h!]
 \centering
\includegraphics[width=0.7\linewidth]{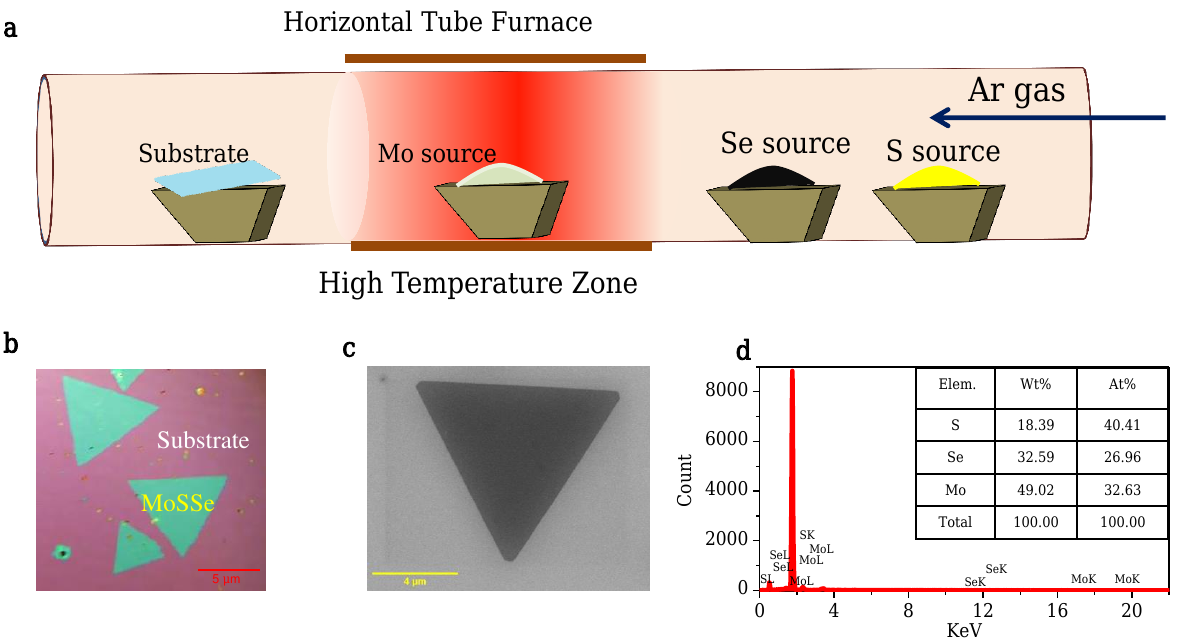}
\caption{{\bf{a}} Schematic representation of 2D TMDCA preparation process using Chemical Vapor Deposition (CVD) technique. {\bf{b}} and {\bf{c}} Optical and SEM images of as-grown TMDCA flake. {\bf{d}} EDX spectrum and elementary percentage of Mo, S and Se of our prepared TMDCA sample is given in inset table.}
\label{Supplementary Figure 1.}
\end{figure*}

\subsection*{Supplementary Note 3: Characterizations of TMDCA device}
\begin{figure*} [h!]
 \centering
\includegraphics[width=0.7\linewidth]{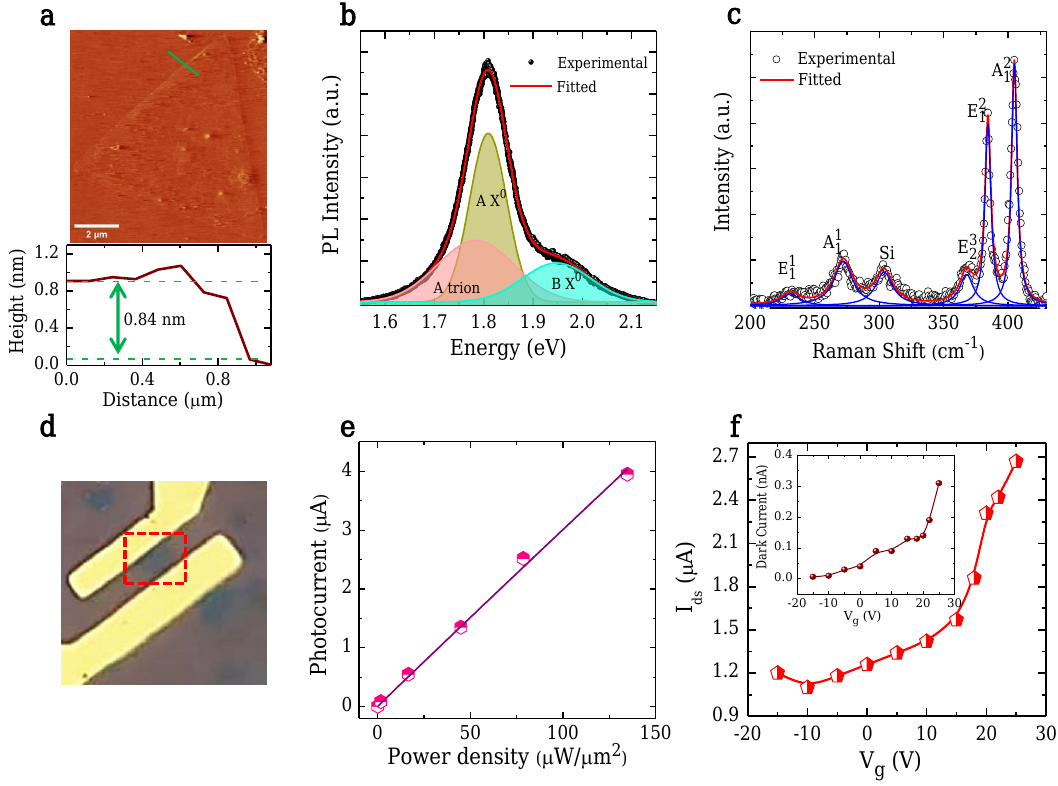}
\caption{ {\bf{a}} Top panel shows AFM image of as-synthesized TMDCA and bottom panel presents height profile from AFM analysis. {\bf{b}} and {\bf{c}} PL and Raman spectra of the TMDCA, respectively. {\bf{d}} Optical image of as fabricated device (TMDCA sample is marked by the dotted red coloured rectangle). {\bf{e}} Photocurrent dependence on the light excitation power density. and {\bf{f}} Transfer characteristics of the device under dark (inset) and illumination conditions with $V_{ds}$= 2 V.}
\label{Supplementary Figure 1.}
\end{figure*}
\newpage
\subsection*{Supplementary Note 4: Piezoelectric force microscopy (PFM)}
Piezoelectric force microscopy (PFM) is a powerful technique for assessing piezoelectric phenomenon at the nanoscale level. We conducted measurements in the vertical mode on SiO$_2$/Si substrate as illustrated schematically in Supplementary Fig. 9. In this method, the conductive tip oscillates in contact with the sample surface, and the detected signal represents the first harmonic component, reflecting the vertical out-of-plane deformation. The piezoresponse amplitude (A) can be described as \cite{LuM, XieS}:
$A = d_{33}V_{d}Q$.
Here, $d_{33}$ represents the piezoelectric coefficient, $V_d$ is the AC driving voltage, and $Q$ is the resonance quality factor. Thus, the deflection amplitude is determined by the piezoelectric coefficient, while the piezoelectric phase provides information about the polarization direction. In our experiment, the cantilevers used had spring constants of approximately 2.8 N/m and free-air resonances at 75 kHz. The Si probes were coated with Ti/Ir for electrical conductivity. The resonance frequency and quality factor are around 320 kHz and 100 respectively. 
\begin{figure*}[h!]
 \centering
\includegraphics[width=0.4\linewidth]{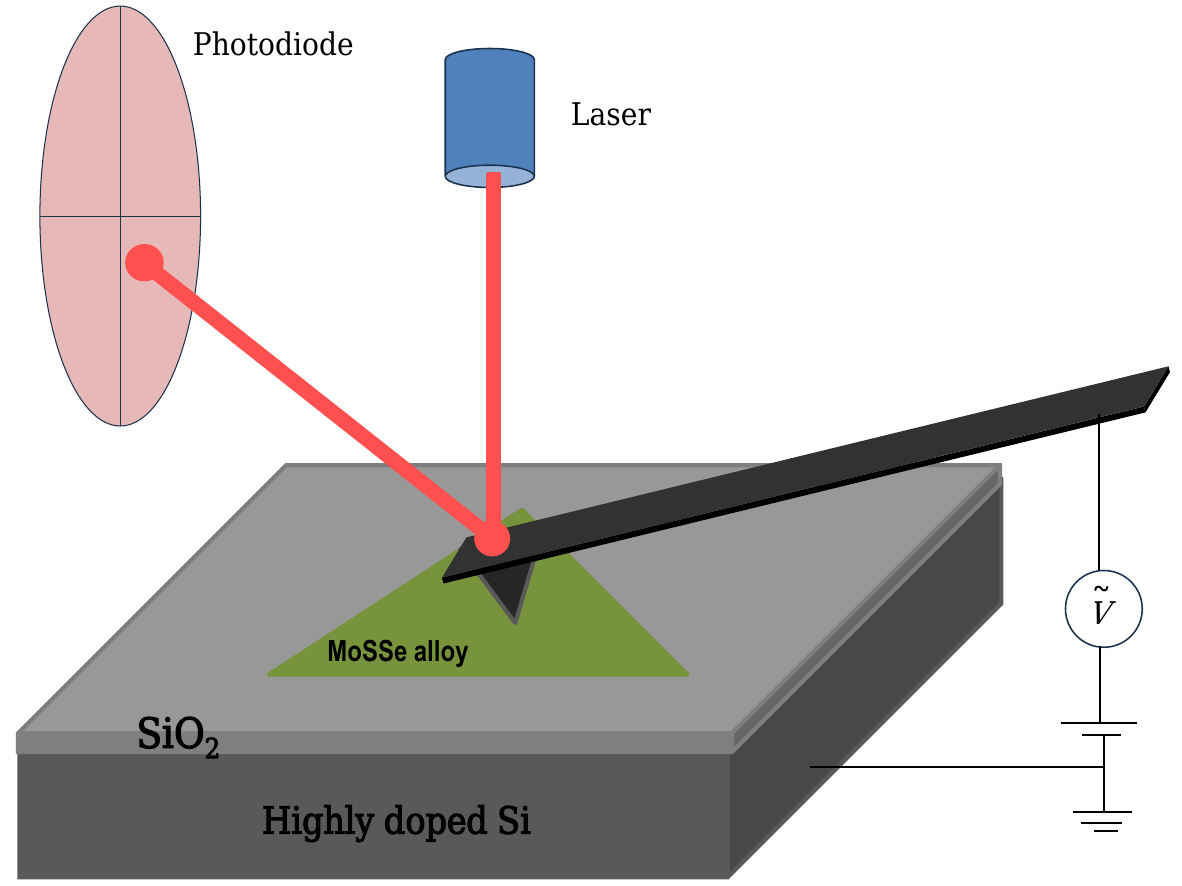}
\caption{Schematic representation of PFM measurement on TMDCA. }
\label{Supplementary Figure 1.}
\end{figure*}
\begin{figure*} [b!]
 \centering
\includegraphics[width=0.65\linewidth]{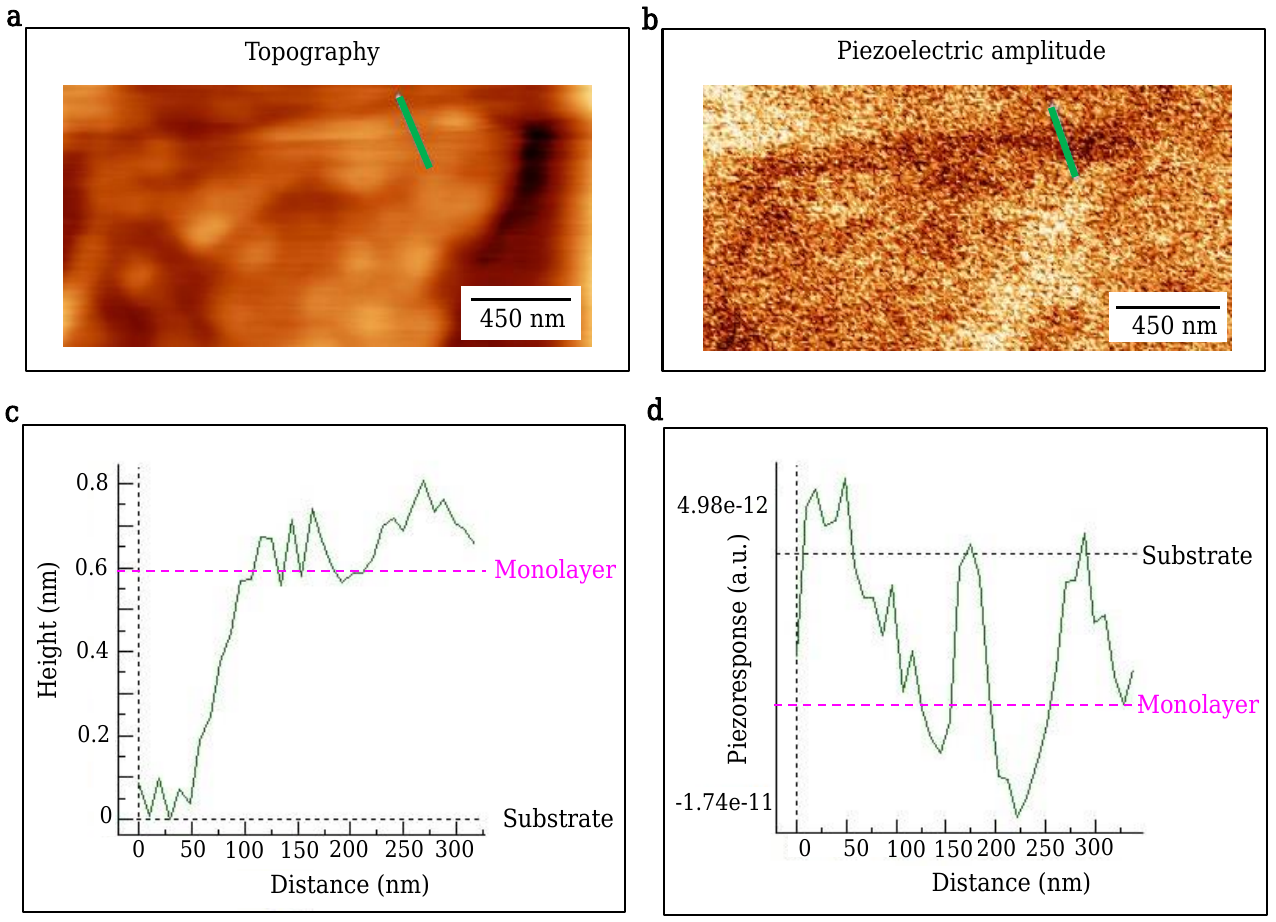}
\caption{PFM measurement on monolayer TMDCA.  {\bf{a}} Topography  {\bf{b}} Piezoelectric amplitude. For parallel comparison, cross-sections along the green lines in both images are compared. {\bf{c}} Height and {\bf{d}} Piezoelectric amplitude line profile of monolayer TMDCA. }
\label{Supplementary Figure 1.}
\end{figure*}
\begin{figure*} [t!]
 \centering
\includegraphics[width=0.6\linewidth]{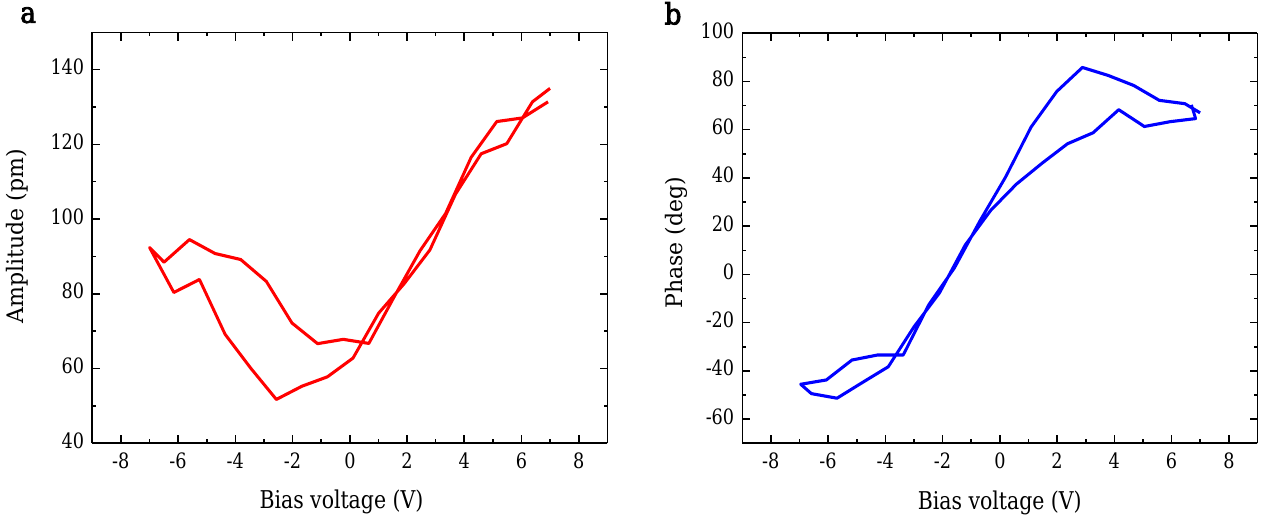}
\caption{ PFM measurement on monolayer TMDCA.{\bf{a}} and {\bf{b}}  The PFM amplitude and phase as a function of  applied bias from the monolayer sample.}
\label{Supplementary Figure 1.}
\end{figure*}
\begin{figure*} [h!]
 \centering
\includegraphics[width=0.65\linewidth]{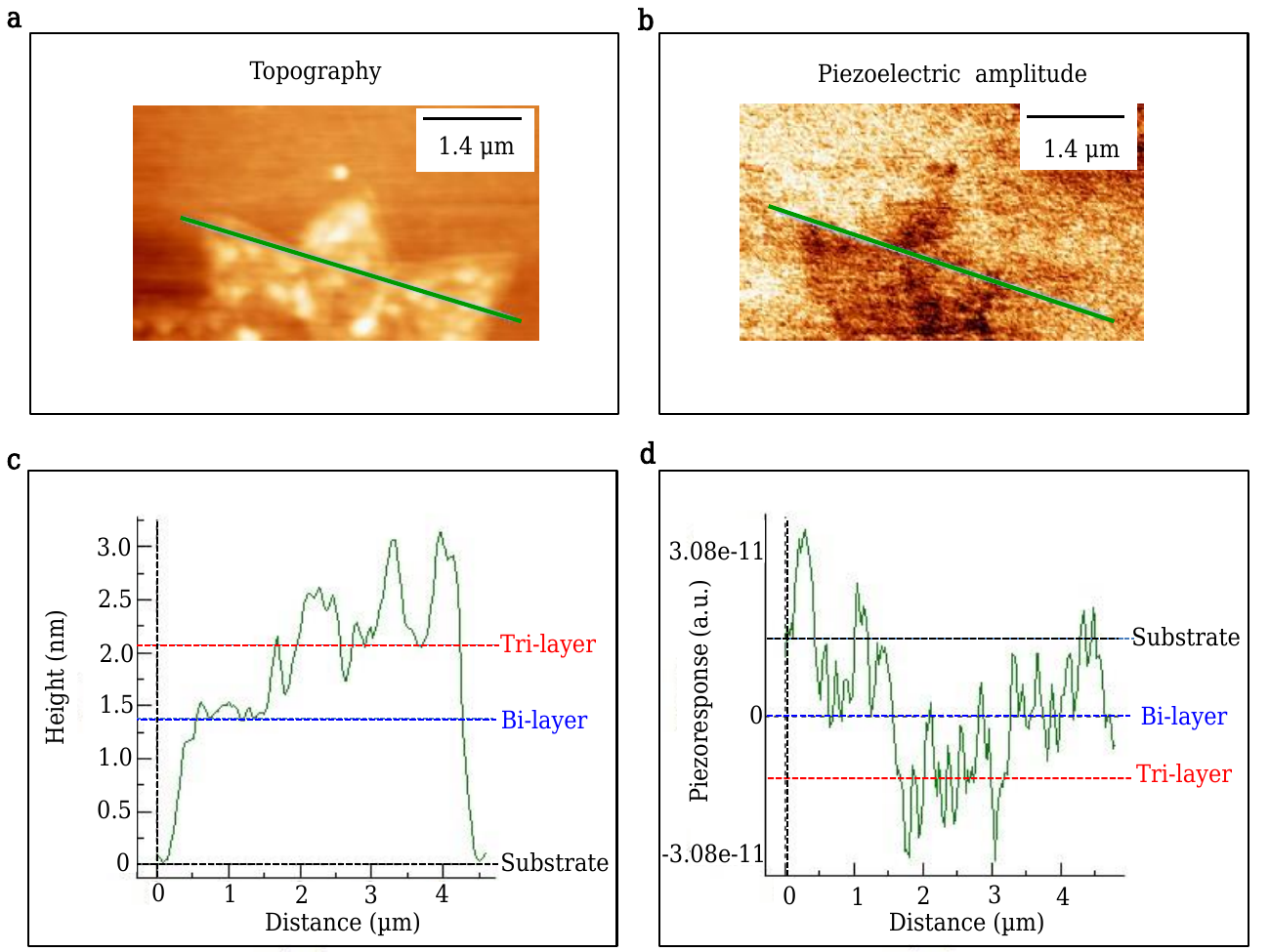}
\caption{ PFM measurement on bilayer and trilayer TMDCA.  {\bf{a}} Topography, {\bf{b}} Piezoelectric amplitude. For parallel comparison, cross-sections along the green lines in both images are compared. {\bf{c}} Height and {\bf{d}} Piezoelectric amplitude line profile of bilayer and trilayer TMDCA.}
\label{Supplementary Figure 1.}
\end{figure*}
\begin{figure*} [b!]
 \centering
\includegraphics[width=0.6\linewidth]{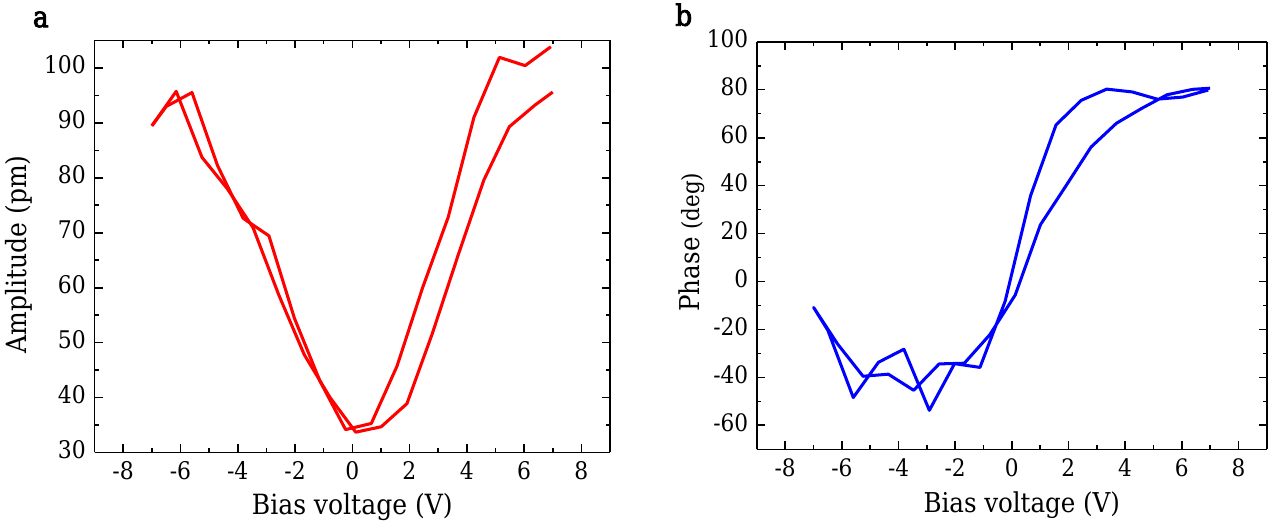}
\caption{ PFM measurement on bilayer and trilayer TMDCA. {\bf{a}} and {\bf{b}} The PFM amplitude and phase as a function of  applied bias from the monolayer sample.}
\label{Supplementary Figure 1.}
\end{figure*}

\begin{figure*} [h!]
 \centering
\includegraphics[width=0.7\linewidth]{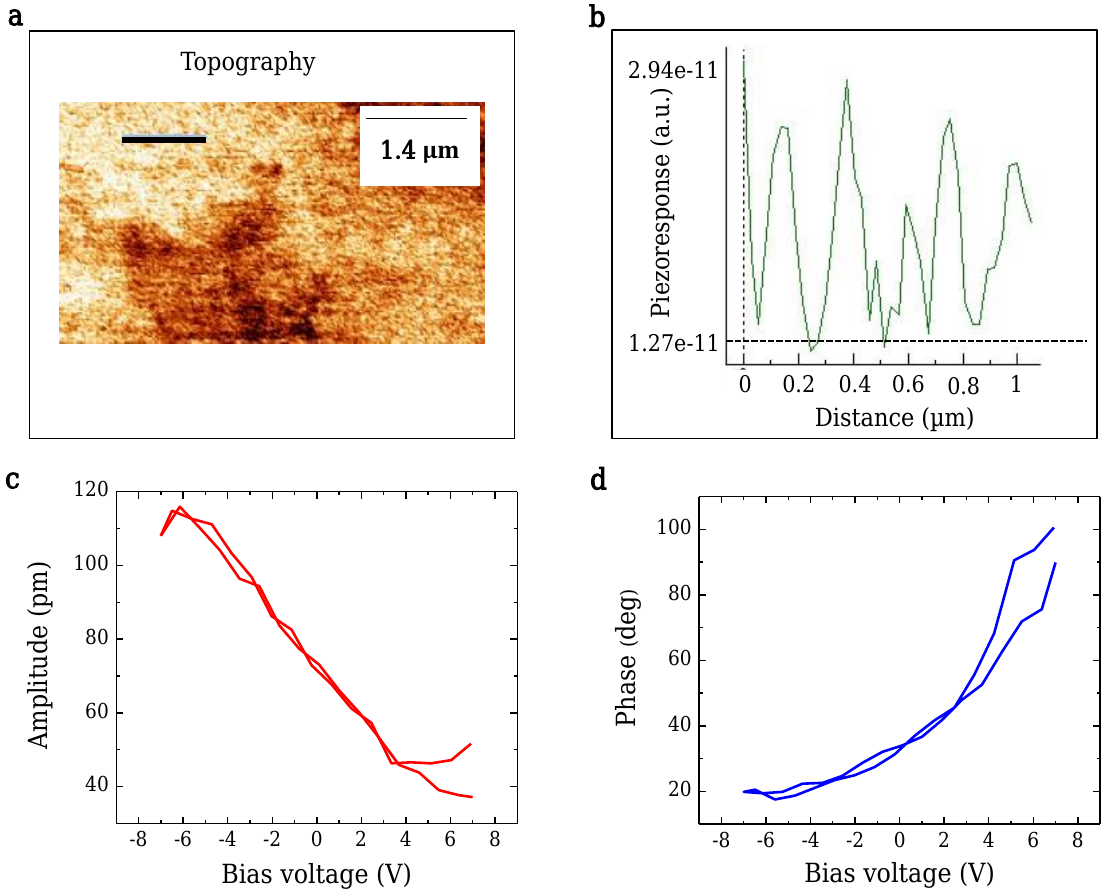}
\caption{ PFM measurement on SiO$_2$/Si substrate.  {\bf{a}}  Topography {\bf{b}} Piezoelectric amplitude line profile . {\bf{c}} and {\bf{d}}  The PFM amplitude and phase as a function of  applied bias from the substrate.}
\label{Supplementary Figure 1.}
\end{figure*}
\newpage

\subsection*{Supplementary Note 5: $|C|$, $|L_1|$, $|L_2|$ and $F$ values with angle of incidence of illumination under 1.96 eV excitation.}
\begin{figure*} [h!]
 \centering
\includegraphics[width=0.65\linewidth]{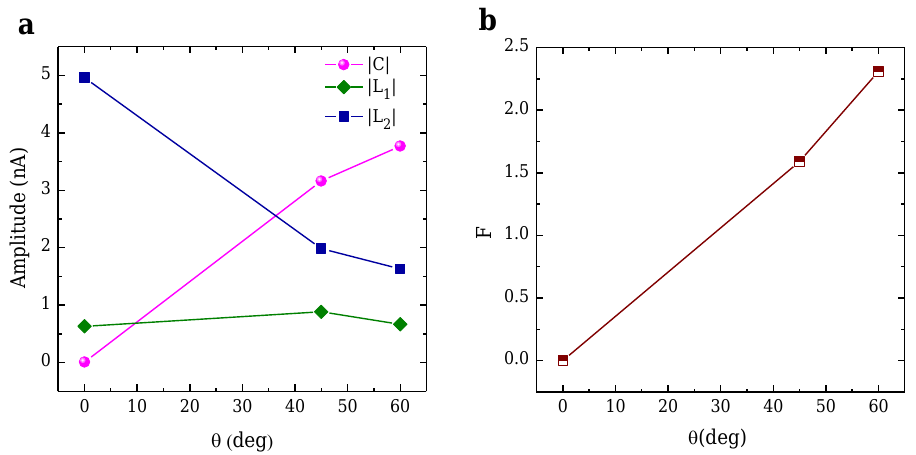}
\caption{{\bf{a}} and {\bf{b}} Plot of extracted $|C|$, $|L_1|$ and $|L_2|$ values using Eq. 3 and F with angle of incidence of illumination ($\theta$) under 1.96 eV excitation.}
\label{Supplementary Figure 1.}
\end{figure*}
\newpage
\subsection*{Supplementary Note 6: $|L_1|$, $|L_2|$ and $F$ values with gate voltage under 1.96 eV excitation}
\begin{figure*} [h!]
 \centering
\includegraphics[width=0.65\linewidth]{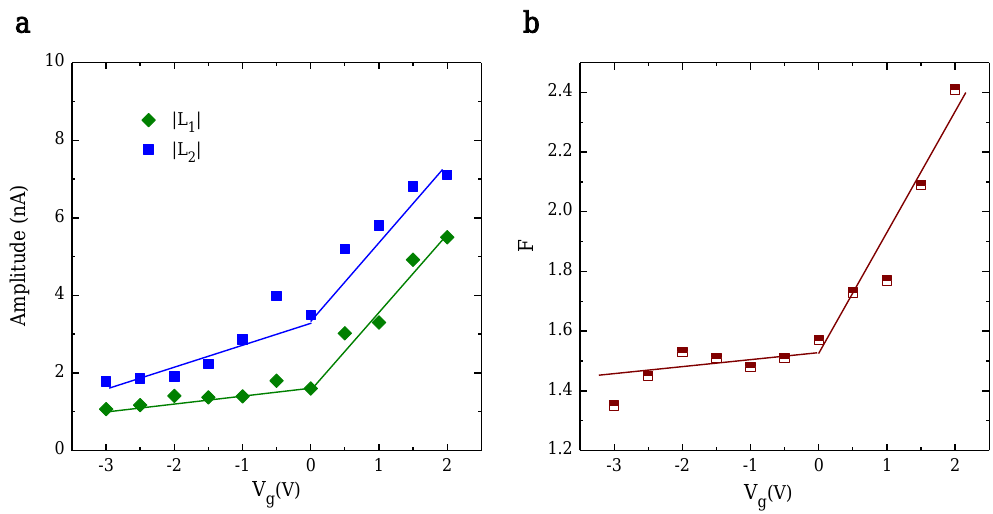}
\caption{ {\bf{a}} Plot of extracted $|L_1|$ and $|L_2|$ values from gate dependent photocurrent data using Eq. 3 and {\bf{b}} F with gate voltage under 1.96 eV excitation.}
\label{Supplementary Figure 1}
\end{figure*}
\subsection*{Supplementary Note 7: Helicity-dependent  Photocurrent measurements on two additional devices}
\begin{figure*}[h!]
 \centering
\includegraphics[width=0.7\linewidth]{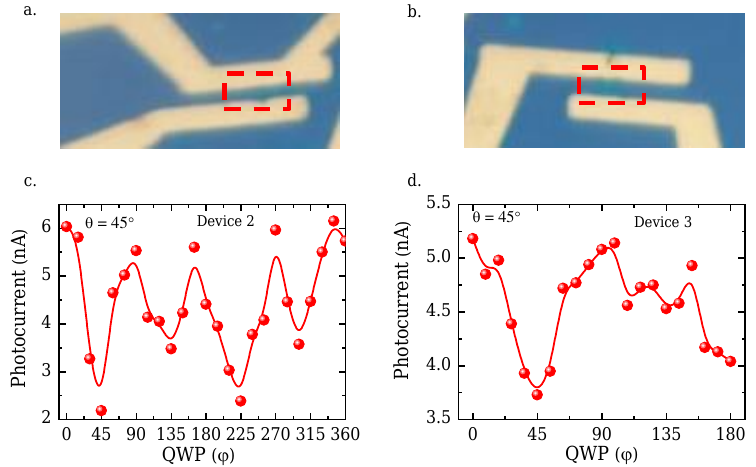}
\caption{ {\bf{a}} and {\bf{b}} Optical images of two additional devices 
(TMDCA sample is marked by the dotted red coloured rectangle) and their corresponding  helicity-dependent photocurrent data using excitation energy 1.96 eV {\bf{c}} and {\bf{d}}.}
\label{Supplementary Figure 1}
\end{figure*}
\newpage
\subsection*{Supplementary Note 8: PL study of few layer and bulk TMDCA.}
\begin{figure*} [h!]
 \centering
\includegraphics[width=0.4\linewidth]{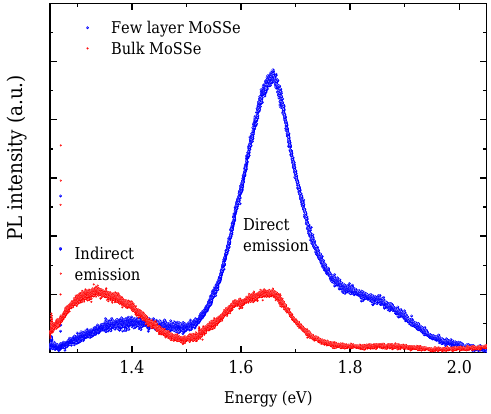}
\caption{PL spectra of few layer and bulk TMDCA under laser excitation of 2.54 eV energy.}
\label{Supplementary Figure 1.}
\end{figure*}
\subsection*{Supplementary Note 9: Helicity-dependent PL study under 2.54 eV excitation energy}
\begin{figure*}[h!]
 \centering
\includegraphics[width=0.45\linewidth]{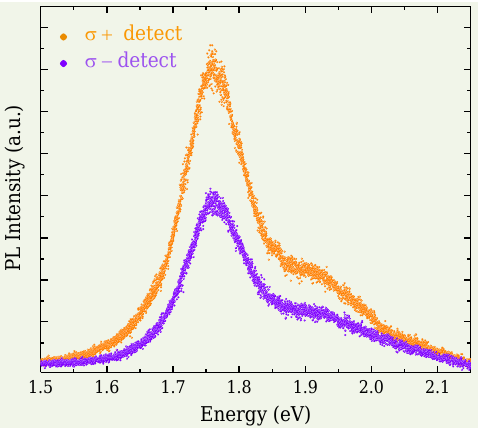}
\caption{Helicity-dependent PL spectra under 2.54 eV excitation energy.}
\label{Supplementary Figure 1.}
\end{figure*}
\newpage
\subsection*{Supplementary Note 10: Helicity-dependent  Photocurrent measurements under 2.54 eV excitation}
\begin{figure*}[h!]
 \centering
\includegraphics[width=0.5\linewidth]{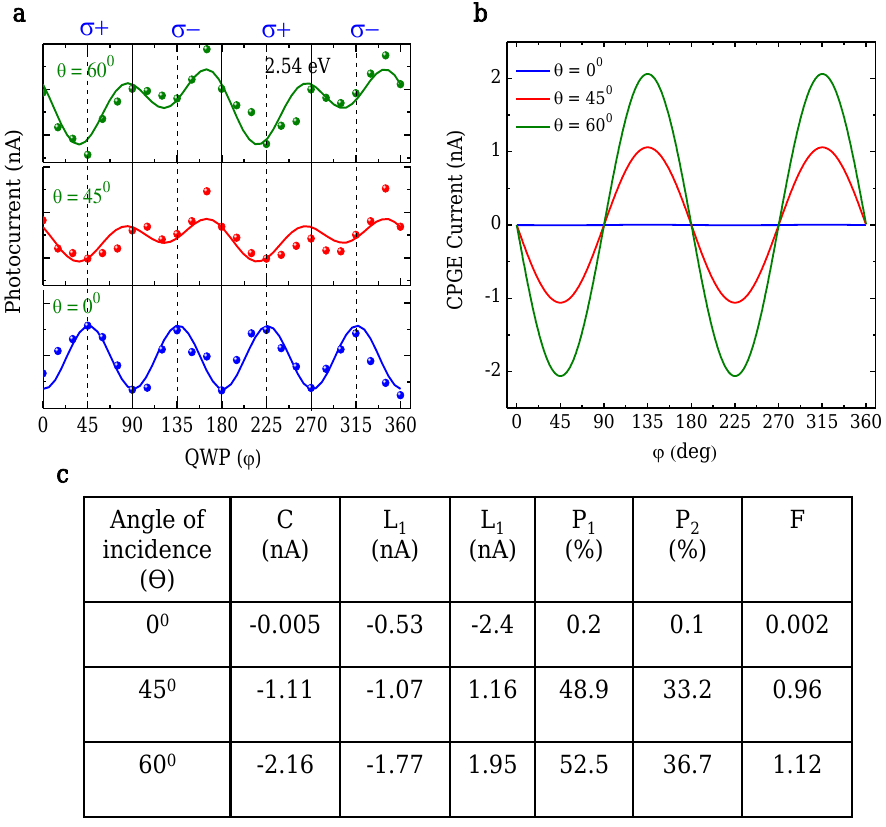}
\caption{ {\bf{a}} Helicity-dependent photocurrent measurement using excitation energy 2.54 eV, where scattered points are the experimental data and solid lines represent the fitted data using Eq. 3. {\bf{b}} Plot of CPGE current as a function of polarization-angle $\varphi$ at varying incidence angles of illumination and {\bf{c}} Value of fitted parameters ($|C|$, $|L_1|$, $|L_2|$), P$_1$, P$_2$ and F for all $\theta$ under 2.54 eV energy excitation.}
\label{Supplementary Figure 1.}
\end{figure*}
\subsection*{Supplementary Note 11: Helicity-dependent photocurrent measurements of bilayer TMDCA}
\begin{figure*}[h!]
 \centering
\includegraphics[width=0.6\linewidth]{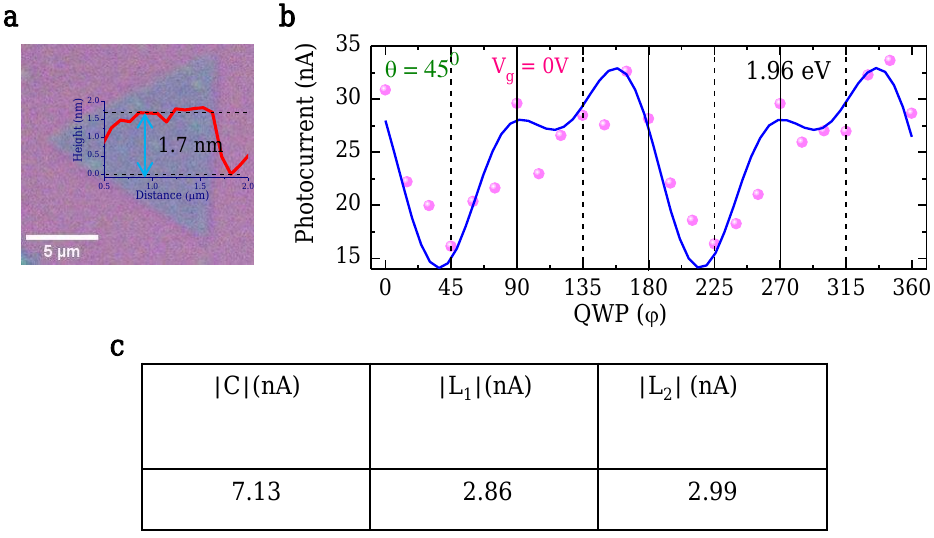}
\caption{{\bf{a}} Optical image of bilayer TMDCA with height profile. {\bf{b}} Helicity-dependent photocurrent measurement using laser excitation energy 1.96 eV, where scattered points are the experimental data and solid lines represent the fitted data using Eq. 3. {\bf{c}} Fitted parameters ($|C|$, $|L_1|$, $|L_2|$) for bilayer device.}
\label{Supplementary Figure 1.}
\end{figure*}

\newpage
\def\bibsection{\subsection*{\refname}}

\subsection*{Acknowledgements} 
C. Nayak acknowledges the INSPIRE Fellowship Programme,
DST, Government of India, for granting her a research
fellowship with Registration Number IF180057. S. Masanta expresses gratitude to the Council of Scientific $\&$ Industrial Research (CSIR), India, for the financial support provided through the NET-SRF award (File Number 09/015(0531)/2018-EMRI). S. Ghosh acknowledges the computational resources provided by the Swedish National Infrastructure for Computing (SNIC) at NSC, PDC, and HPC2N partially funded by the Swedish Research Council (Grant No. 2018-05973). A. N. Pal acknowledges DST Nano Mission: DST/NM/TUE/QM-10/2019. I. Bose acknowledges the support of NASI, Allahabad, India under the Honorary Scientist Scheme. A. Singha acknowledges financial support from the Science and Engineering Research Board (SERB), India (File Number EMR/2017/002107). 
\subsection*{Author contributions}
C. Nayak, and A. Singha conceived and designed the experiments. C. Nayak and S. Masanta synthesized the 2D TMDCA and performed the experiments. C. Nayak, S. Masanta and A. Singha analysed the data. S. Moulick and A. N. Pal fabricated the devices and participated in AFM study. S. Ghosh performed first principles DFT calculations. I. Bose performed the theoretical model calculations and contributed towards conceptualisation and physical interpretations. C. Nayak, S. Ghosh, I. Bose and A. Singha contributed to the writing of the manuscript. 

\end{document}